\documentclass[12pt,epsfig]{article}
\usepackage{amsfonts}
\usepackage{amsmath}
\usepackage{amssymb}
\usepackage{color}

\newcommand{\rd}{\textcolor{red}}

\def\*{{\phantom *}}

\newtheorem{postulate}{Hypothesis}[section]  
\newcommand{\bhypo}{\begin{postulate} \it}
\newcommand{\ehypo}{\end{postulate}}

\newtheorem{theorem}{Theorem}[section]  
\newcommand{\btheo}{\begin{theorem} \it}
\newcommand{\etheo}{\end{theorem}}

\newtheorem{proposition}{Proposition}[section]  
\newcommand{\bprop}{\begin{proposition}\it}
\newcommand{\eprop}{\end{proposition}}

\newtheorem{corollary}{Corollary}[section]  
\newcommand{\bcorol}{\begin{corollary} \it}
\newcommand{\ecorol}{\end{corollary}}

\newtheorem{lemma}{Lemma}[section]  
\newcommand{\blem}{\begin{lemma}\it}
\newcommand{\elem}{\end{lemma}}

\newtheorem{remark}{Remark}[section]   
\def\brem{\begin{remark} \rm }
\def\erem{\end{remark}}

\newtheorem{naming}{Definition}[section]   
\newcommand{\bdefi}{\begin{naming} \rm }
\newcommand{\edefi}{\end{naming}}

\begin{document}
\qquad \hfill \rd{\textbf{DRAFT April 26, 2009}}

\phantom{.}
\vskip 1.5cm
\begin{center}
{\Large{\bf {Scaling approach to existence of long cycles in Casimir boxes}}}
\linebreak
\vskip 1cm

\setcounter{footnote}{0}
\renewcommand{\thefootnote}{\arabic{footnote}}

\vspace{1cm}

\textbf{Mathieu BEAU}\footnote{E-mail: mathieubeau@hotmail.fr}

\vspace{0.2 cm}

\textbf{Universit\'e de la M\'editerran\'ee and Centre de Physique
Th\'eorique - UMR 6207 \footnote{Universit\'{e} de Provence - Aix-Marseille I,
Universit\'{e} de la M\'{e}diterran\'{e}e - Aix-Marseille II,
Universit\'{e} du Sud - Toulon - Var, FRUMAM (FR 2291)}, \\ Luminy - Case 907,
13288 Marseille, Cedex 09, France}

\vspace{0.7 cm}

\vspace{1.5cm}

\end{center}
\begin{abstract}
{We analyse the concept of generalized Bose-Einstein condensation (g-BEC), known since 1982 for the perfect Bose gas (PBG)
in the Casimir (or anisotropic) boxes. Our aim is to establish a relation between this phenomenon and two
concepts:  the concept of long cycles  and the Off-Diagonal-Long-Range-Order (ODLRO), which are usually considered as some
adequate way to describe the standard BEC on the ground state for the cubic boxes.} First we show that these three
criterions are equivalent in this {latter} case. Then, basing on a scaling approach, we revise formulation of these
concepts to prove that the classification of the g-BEC into three types I,II,III, implies a hierarchy of long cycles
(depending on their {size scale}) as well as a hierarchy of ODLRO which depends on the coherence length of the condensate.
\end{abstract}

\bigskip

\noindent \textbf{Keywords:} Generalized Bose-Einstein condensation, (Long) Cycles,
Off-Diagonal-Long-Range-Order (ODLRO), perfect Bose gas, Casimir boxes, Scaling.

\bigskip

\noindent \textbf{PACS:} 05.30.Jp, 03.75.Hh   \\

\newpage
\tableofcontents
\section{Introduction}
\setcounter{equation}{0}
\renewcommand{\theequation}{\arabic{section}.\arabic{equation}} 
\noindent\emph{About the Bose-Einstein condensation} \\

The Bose-Einstein condensation {(BEC)} {predicted in 1925 was discovered first} in superfluid ${^4}He$ in 1975
{in deep-inelastic neutron scattering experiences,}
see \cite{Z-B} for historical remarks. {Since 1995} it attracts a lot of attention of theoretical and mathematical
physicists motivated by experiments with ultra-cold gases in traps  \cite{L-S-S-Y}.

Recently experimentalists discovered {some new peculiarities} of the Bose-Einstein condensation in very anisotropic traps
\cite{M-H-U-B}. This may to imply in certain cases a {(what they called)} \textit{fragmentation} of the condensate.
In fact this phenomenon was predicted long time ago by H.Casimir \cite{C}. Then it was carefully studied by  mathematical
physicists and now it is known under the name of the \textit{generalized} Bose-Einstein condensation
(g-BEC) \textit{\`{a} la van den Berg-Lewis-Pul\'{e}} \cite{Z-B}. After the first publication by M.van den Berg and J.Lewis
in 1982 \cite{vdB-L}, a set of articles treated different cases of the Casimir's anisotropic box, boundary conditions and
external potentials, \cite{vdB}, \cite{vdB-L}, \cite{vdB-L-P}, \cite{vdB-L-L} {has appeared.}
They classified the Bose-Einstein condensation into three types. If a finite/infinite number of one-particle
{kinetic-energy} quantum states are macroscopically occupied, then the Bose gas manifests g-BEC of type I/II.
If there are no states macroscopically occupied, although in the thermodynamical limit
a macroscopic number of particles is accumulated in the ground (zero-mode) state, then the Bose gas manifests g-BEC of
the type III.

It is worthy to note that not only {the box anisotropy or boundary conditions}, but also the \textit{interaction} between
particles is able to modify the type of the condensate, see \cite{M-V} and \cite{Br-Z}.

In 1953 R.Feynman \cite{F} introduced a concept of \textit{cycles} by rewriting the partition function of the
boson gas using the Bose-statistic. In the Feynman-Kac representation, cycles are {closed random Wiener trajectories.}
This was a mathematical framework for the path-integral formulation of the quantum statistical mechanics.
Heuristically at low temperature (or at high density), bosons {start to form} a cycles of different sizes and
below {certain} critical point, the \textit{infinite} cycles appear in the Feynman-Kac representation. {Since
this Feynman observation,} it is a common wisdom to count this equivalent to the (type I) Bose-Einstein condensation
\cite{P-O}. For the perfect Bose gas it was proven in \cite{S} and for some weakly interacting Bose gases only recently
in \cite{D-M-P} and \cite{U}.

In 1956, O.Penrose and L.Onsager \cite{P-O} introduced, via a reduced density matrix,
the concept of the Off-Diagonal-Long-Range-Order (ODLRO) to give an alternative description of the Bose-Einstein
condensation in ideal or interacting Bose-gases. They worked out a convincing arguments that existence of
the ODLRO is equivalent to the Bose-Einstein condensation.
In fact the ODLRO measures in boson systems a correlation between two infinitely distant points.
Recently in papers \cite{U} and \cite{U2} a contact between the ODLRO and the existence of the long cycles was studied.\\

\noindent\emph{Conceptual and physical problems}\\

{Although the ODLRO} is usually considered as a criterion of (type I) BEC, it is not evident that
the ODLRO is equivalent to the generalized BEC.
Moreover, it is not evident that the presence of long cycles is equivalent to generalized BEC.
Therefore, {our} paper is motivated by two  questions:

Are the different criteria of Bose-Einstein condensation (generalized BEC, long cycles and ODLRO)
equivalent for the \textit{perfect} Bose-gas in anisotropic (Casimir) boxes (Section 2) ?
Hence, the purpose of the Section 2 is an extension of known results on the long cycles, see
A.Suto \cite{S}, T.Dorlas, P.Martin and J.Pul\'e \cite{D-M-P}, D.Ueltschi \cite{U},
to the Casimir boxes, where the concept of the generalized Bose-Einstein condensation can be explicitly verified.

The second question is how we can classify different types of generalized Bose-Einstein condensation with the
help of the concept of cycles and {with help of} the ODLRO (Sections 3, 4 and 5)?

The aim of this paper is {to relate the van den-Berg-Lewis-Pul\'{e} classification of g-BEC (type I, II and III)
in anisotropic boxes with} a hierarchy of long cycles and {with} the corresponding hierarchy of the ODLRO.
More precisely, we would like to know, what is the scale of different sizes of the long cycles (macroscopic or not)
and correlations (the coherence length) of the condensate ?

Our arguments are based on the \textit{scaling approach}. To this end, we propose a scaling formulation for
the condensate density and for the notion of long cycles (Section 3), as well as for the reduced density
matrix and the ODLRO (Section 5).

Since the seventies \cite{dG} scaling concepts {are also} used in polymer physics.
In the present paper we adapt this scaling approach to the cycles (and two-point correlation function)
{in the both cases} because there is a deep analogy between cycle representation of boson systems and polymers \cite{C-S}.
Below we give a mathematical presentation of this scaling concept concerning the Bose-Einstein condensation
of the Ideal Bose-Gas in Casimir boxes.

{This concept} is relevant {in the physics of quantum coherent states}, since it relates the box geometry constrains
to the coherence shape of condensate clouds and to the "geometry" of the boson cycles ({polymers' shape}).
Heuristically, there is a scaling relation between the \textit{coherence length} $r$ and the size of long cycles
involved $n$ bosons, which is given by $r=\lambda_{\beta}n^{1/2}${, provided correlation function has the form
(\ref{RkODLRO}). Here $\lambda_{\beta}=\hbar\sqrt{2\pi\beta/m}$ is the thermal de Broglie length}.
This relation is analogous to the scaling law for the \textit{ideal polymer chain} \cite{dG}, where the size of the chain
$R$ is proportional to the number of monomers $N$: $R=l_{0}N^{1/2}$, where $l_{0}$ is the effective length of a monomer. \\

\emph{Results} \\

This paper contains two kind of results interesting for physical properties of the Bose-gas in Casimir boxes:
$\Lambda=L_{1}\times L_{2}\times L_{3}$,
with $L_{\nu}=V^{\alpha_{\nu}},\ \nu=1,2,3,$, where $\alpha_{1}+\alpha_{2}+\alpha_{3}=1$.
These results establish a contact between generalized van den Berg-Lewis-Pul\'{e} condensate and the experimental data
concerning the {BEC fragmentation}  \cite{M-H-U-B}.

The first result {states that} the number of particles $N_{0}$ in the condensate for finite-size system ($N$ particles)
is (see section 3.2):
\begin{eqnarray*}
N_{0}=n_{1}+n_{2}+...+n_{M},
\end{eqnarray*}
where $n_{i}$ are the numbers of particles in the condensate states.
Here $n_{i}=O(N)$ and $M=O(1)$, if $\alpha_{1}\leqslant1/2$ (see Theorem \ref{thm1} and \ref{thm2})
or $n_{i}=O(N^{\delta})$ and $M=O(N^{1-\delta})$ (such that $N_{0}=O(N)$), $\delta=2(1-\alpha_{1})<1$, if $\alpha_{1}>1/2$
(see Theorem \ref{thm3}). This result {has a direct relation} to the fragmentation theory of Bose-Einstein condensation
\cite{M-H-U-B}. The second part of this result (see Section 4) is that the order of the size of long cycles
is \textit{macroscopic} (i.e. of the order $O(N)$), if we have generalized BEC of type I or II (when $M=O(N)$)
and that the order of the size of the {long} cycles is \textit{microscopic} (of the order $O(N^{\delta})$)
for the generalized BEC of type III, see Theorem \ref{thm1}, \ref{thm2} and \ref{thm3}.

The second result is that by virtue of Theorem \ref{TmODLRO} the two points correlation functions
(at different scales) have the form:
\begin{eqnarray*}
\lim_{V\uparrow\infty}\sigma_{\Lambda}(x-x')&=&\rho_{0}(\beta),\ \mathrm{for}\ \|x-x'\|=O(V^{\alpha_{1}}) ,\
\mathrm{if}\ \alpha_{1}<1/2,
{}\\{}&=&\sum_{n_{1}\in\mathbb{Z}^{1}}\frac{\cos(2\pi n y)}{\pi\lambda_{\beta}^{2} n^2 +B}
,\ \mathrm{for}\ \|x-x'\|=yV^{\alpha_{1}} ,\ \mathrm{if}\ \alpha_{1}=1/2,
{}\\{}&=&\rho_{0}(\beta)e^{-2y \sqrt{\pi C}/\lambda_{\beta}}
,\ \mathrm{for}\ \|x-x'\|=yV^{\delta/2} ,\ \mathrm{if}\ \alpha_{1}>1/2,
\end{eqnarray*}
where $\rho_{0}(\beta)$ is the {particle density} in the condensate, $B$ and $C$ are two positive constants
respectively given by (\ref{EqB}) and (\ref{EqC}).
{These formes imply} that the order of the \textit{condensate coherence length} coincides with the
size of the box in the case of g-BEC of type I or II (i.e. long cycles of \textit{macroscopic} size),
but it is smaller if we have generalized BEC of type III (i.e. long cycles of \textit{microscopic} size).
The last case shows decreasing of the coherence length for the elongated condensate.
Notice that existence of this phenomenon is indicated in physical literature \cite{P-S-W}.

To make a contact with experiments involved the cold atoms confined in magnetic traps one should extend these
results to the Bose-gas in a weak harmonic potential. So, an important {task} is the theoretical and experimental
study of very anisotropic {cases} (like quasi-2D or quasi-1D systems) to understand the coherence properties
of condensate, which mimics the Casimir case.

\section{BEC in Casimir boxes}
\setcounter{equation}{0}
\renewcommand{\theequation}{\arabic{section}.\arabic{equation}} 

In this section we give a {short} review of the different concepts {concerning} the Bose-Einstein condensation:
such as occupation number, generalized BEC (g-BEC), cycles, long cycles, reduce density matrix,
Off-Diagonal-Long-Range-Order (ODLRO) and the link between the different criterions of BEC (g-BEC, long cycles and ODLRO).

\subsection{About the concept of g-BEC condensation and the ODLRO}

\noindent\emph{Basic notions}

Let us consider the grand-canonical $(\beta,\mu)$ \textit{perfect} Bose-gas (PBG) in Casimir boxes
$\Lambda=L_{1}\times L_{2}\times L_{3}\in\mathbb{R}^{d=3}$,
of volume $|\Lambda|=V$ with sides length $L_{\nu}=V^{\alpha_{\nu}},\
\nu=1,2,3$, where $\alpha_{1}\geqslant\alpha_{2}\geqslant\alpha_{3}>0,\
\alpha_{1}+\alpha_{2}+\alpha_{3}=1$.
For the single-particle hamiltonian $H_{\Lambda}^{(N=1)}=T_{\Lambda}^{(1)}:=-(\hbar^{2}/2m)\Delta$
with periodic boundary conditions,
we get the dual vector-spaces $\Lambda^{*}$ defined by:
\begin{equation}\label{dual vector-spaces}
\Lambda^{*}=\left\{k\in\mathbb{R}^{3}:k=(\frac{2\pi n_{1}}{V^{\alpha_{1}}},\frac{2\pi n_{2}}{V^{\alpha_{2}}},
\frac{2\pi n_{3}}{V^{\alpha_{3}}});\ n_{\nu}\in\mathbb{Z}^{1}\right\}.
\end{equation}

In the grand-canonical ensemble, the mean values of the $k$-mode particle densities
$\{\rho_{\Lambda}(k)\}_{k\in\Lambda^{*}}$ are:
\begin{equation}\label{meanparticledensities}
\rho_{\Lambda}(k):=\frac{1}{V}\langle N_{\Lambda}(k)\rangle_{\Lambda}(\beta,\mu)
=\frac{1}{V}\frac{1}{e^{\beta(\epsilon_{\Lambda}(k)-\mu)}-1},
\end{equation}
where $\epsilon_{\Lambda}(k)=\hbar^{2}k^{2}/2m,\ k\in\Lambda^{*}$ are the eigenvalue of the Laplacian
with periodic boundary conditions,
$\langle N_{\Lambda}(k)\rangle_{\Lambda}(\beta,\mu)$ is the Gibbs expectation of the
particles number operator $N_{\Lambda}(k)$ in the mode $k$.
The total density of particles is $\rho_{\Lambda}(\beta,\mu):=\sum_{k\in\Lambda^{*}}\rho_{\Lambda}(k)$

Let us recall that the eigenfunctions $\{\Psi_{\Lambda,k}^{(N=1)}(x)\}_{k\in\Lambda^{*}}$
of the single particle hamiltonian $T_{\Lambda}^{(1)}$ are:
\begin{equation}\label{eigenfunction1}
\Psi_{\Lambda,k}^{(1)}(x)=\frac{1}{\sqrt{V}}e^{ik.x}\ .
\end{equation}

\noindent\emph{London scaling approach and g-BEC}

In fact it was F.London \cite{L} who for the first time used implicitly the \textit{scaling}
approach to solve the controversy between the G.Uhlenbeck mathematical arguments against condensation of
the perfect Bose-gas for hight densities and Einstein's intuitive reasoning in favour of this phenomenon.
His line of reasoning was based on the following observations:
since the explicit formula for the total particle density $\rho_{\Lambda}(\beta,\mu)$ in the
box $\Lambda$ is known only in grand-canonical ensemble $(\beta, \mu)$, to ensure fixed density $\rho$
in this box one has first to solve equation $\rho=\rho_{\Lambda}(\beta,\mu)$ which determine the
corresponding value
of the chemical potential $\overline{\mu}_{\Lambda}:= \overline{\mu}_{\Lambda}(\beta,\rho)$.

Then the van den Berg-Lewis-Pul\'e formulation of the g-BEC concept in Casimir boxes gets the form:
\begin{naming}\label{Defg-BEC}
We say that for the grand-canonical PBG manifests the g-BEC for a fixed total density of particle
$\rho$, if one has:
\begin{equation}
\rho_{0}(\beta, \rho) := \lim_{\epsilon\downarrow 0}\lim_{V\uparrow\infty}
\sum_{\{k\in\Lambda^{*}:\|k\|\leqslant \epsilon\}}\rho_{\Lambda}(k) > 0 \ ,
\end{equation}
where $\rho_{\Lambda}(k)$ are defined by (\ref{meanparticledensities}) for
$\mu=\overline{\mu}_{\Lambda}(\beta,\rho)$.
\end{naming}

This motivates the following classification:

\begin{naming}\label{g-BECvanderBergI,II,III}

$\bullet$ One gets g-BEC of type I if a finite number
of the single-particle states are macroscopically occupied.

$\bullet$ There is g-BEC of type II is an infinite (countable) number
of the single-particle states are macroscopically occupied.

$\bullet$ The g-BEC is called type III, if none of the
single-particle state is macroscopically occupied, but $\rho_{0}(\beta,\rho)>0$.
\end{naming}

As usually one introduces for the PBG the critical density, $\rho_{c}(\beta)$ defined by:
\begin{equation}\label{rhoc}
\rho_{c}(\beta):=\sup_{\mu<0}\lim_{V\uparrow\infty}\rho_{\Lambda}(\beta,\mu)=g_{3/2}(1)/\lambda_{\beta}^{3},
\end{equation}
where $g_{s}(z):=\sum_{j=1}^{\infty}z^{j}/j^{s}$ is related to the Riemann zeta-function $\zeta(s):=g_{s}(1)$.
{Here $\lambda_{\beta}=\hbar\sqrt{2\pi\beta/m}$ is the thermal de Broglie length.
Notice that the critical density does not depend of $\alpha_1, \alpha_2, \alpha_3,$ i.e. on geometry of the boxes.
However it is shown in \cite{vdB} that the second critical density noted $\rho_m(\beta)$ to have eventually macroscopic
occupation of states
could be different than $\rho_c(\beta)$ and depend on geometry (see perspectives in the section Concluding remarks),
but for Casimir boxes these two critical densities are equal.}

The following proposition is due to \cite{vdB-L-P}:

\begin{proposition}\label{thm1g-BEC}
For particles densities $\rho<\rho_{c}(\beta)$
there is no g-BEC of the PBG in Casimir boxes and the chemical potential
$\mu = \overline{\mu}(\beta,\rho)$ is a unique solution of equation:
\begin{equation}
\rho=g_{3/2}(e^{\beta\mu})/\lambda_{\beta}^{3}
\end{equation}

Let $1/2>\alpha_{1}$, then for a fixed particles density
$\rho>\rho_{c}(\beta)$ the chemical potential
$\overline{\mu}_{\Lambda} = -A/\beta V+o(1/V)$, with $A>0$
and there is g-BEC of the type I in the single zero-mode $k=0$.
Here $A$ is a solution of equation :
\begin{equation}\label{EqA}
\rho-\rho_{c}(\beta)=\frac{1}{A}\ .
\end{equation}

If $1/2=\alpha_{1}$, then the chemical potential $\overline{\mu}_{\Lambda} = -B/\beta V+o(1/V)$,
with $B>0$ and one gets the g-BEC of type II in the infinite number of modes:
\begin{eqnarray*}
\lim_{V\uparrow\infty}\rho_{\Lambda}(k)&=&\frac{1}{B+\pi\lambda_{\beta}^{2}n_{1}^{2}},\ \mathrm{for}\
k=(2\pi n_{1}/V^{\alpha_{1}},0,0),\ n_{1}\in\mathbb{Z}^{1}\ ,
{}\\{}&=&0,\ \mathrm{for}\ k\neq(2\pi n_{1}/V^{\alpha_{1}},0,0),\ n_{1}\in\mathbb{Z}^{1}\ .
\end{eqnarray*}
Here $B$ is a solution of equation:
\begin{equation}\label{EqB}
\rho-\rho_{c}(\beta)=\sum_{n_{1}\in\mathbb{Z}^{1}}\frac{1}{B+\pi\lambda_{\beta}^{2}n_{1}^{2}}\ .
\end{equation}

If $1/2<\alpha_{1}$, then the chemical potential $\overline{\mu}_{\Lambda}=-C/\beta V^{\delta}+o(1/V^{\delta})$,
with $\delta=2(1-\alpha_{1})$ and $C>0$. The corresponding g-BEC is of the type III:
for all $k\in\Lambda^{*}$ we have $\lim_{V\uparrow\infty}\rho_{\Lambda}(k)=0$, although
$\rho_{0}(\beta,\rho)>0$ and $C$ is a solution of equation:
\begin{equation}\label{EqC}
\rho-\rho_{c}(\beta)=\frac{\sqrt{\pi}}{\lambda_{\beta}C^{1/2}}\ .
\end{equation}
\end{proposition}


\noindent\emph{Reduced density matrix and ODLRO} \\

\begin{naming}\label{DefODLRO}
We say that the PBG manifests for a fixed total density of particles $\rho$ an ODLRO, if {one} gets
a nontrivial limit:
\begin{equation}
\sigma(\beta,\rho) := \lim_{\|x-x'\|\uparrow\infty}\lim_{V\uparrow\infty}\sigma_{\Lambda}(\beta,\rho;x,x')> 0 \ ,
\end{equation}
where $\sigma(\beta,\rho;x,x'):=\lim_{V\uparrow\infty}\sigma_{\Lambda}(\beta,\rho;x,x')$
is the two-point correlation function, defined by:
\begin{equation}\label{EqDefCorrel}
\sigma_{\Lambda}(\beta,\rho;x,x'):=
\sum_{k\in\Lambda^{*}}\langle N_{\Lambda}(k)\rangle_{\Lambda}(\beta,\rho)
\Psi_{\Lambda,k}^{(1)}(x)\Psi_{\Lambda,k}^{(1)}(x')^{*}=\sum_{k\in\Lambda^{*}}\rho_{\Lambda}(k)e^{ik.(x-x')}\ ,
\end{equation}
for periodic boundary conditions.
Here $\rho_{\Lambda}(k)$ is defined by (\ref{meanparticledensities}).
\end{naming}

Recall that according to Definition \ref{Defg-BEC}
for a non-zero BEC of types I and II we obtain in the thermodynamical limit a nontrivial particle density
in the mode null, i.e. particle density with de Broglie wave-length equals infinity.
This allows a \textit{communication} in the condensate through the whole space (in $\mathbb{R}^{3}$).
{This fact is less evident for the g-BEC of the type III, but some heuristic arguments
show that the g-BEC and the ODLRO are equivalent for the PBG in Casimir boxes:}
\begin{theorem}\label{ThmODLRO}
Consider the PBG in Casimir boxes $\Lambda$. Then:
\begin{eqnarray}
\sigma(\beta,\rho)&=&0,\ \mathrm{for}\ \rho<\rho_{c},
{}\nonumber\\{}&=&\rho-\rho_{c}(\beta),\ \mathrm{for}\ \rho>\rho_{c},
\end{eqnarray}
i.e. the ODLRO is non-zero if and only if there is g-BEC.
\end{theorem}

\noindent\emph{Proof} \\

We split the correlation function in two parts,
the first one is the correlation due to the (future) condensate
and the second one corresponds to the particles outside the condensate:
\begin{eqnarray*}
\sigma(\beta,\rho;x,x')&:=&\lim_{\epsilon\downarrow0}\lim_{V\uparrow\infty}
\sum_{k\in\{\Lambda^{*}:\|k\|\leqslant\epsilon\}}\rho_{\Lambda}(k)e^{ik.(x-x')}
+\lim_{\epsilon\downarrow0}\lim_{V\uparrow\infty}\sum_{k\in\{\Lambda^{*}:\|k\|>\epsilon\}}
\rho_{\Lambda}(k)e^{ik.(x-x')}{}\\{}&=&
\lim_{\epsilon\downarrow0}\lim_{V\uparrow\infty}\sum_{k\in\{\Lambda^{*}:\|k\|\leqslant\epsilon\}}\rho_{\Lambda}(k)
+\frac{1}{(2\pi)^{3}}\int_{k\in\mathbb{R}^{3}}dk
\frac{e^{ik.(x-x')}}{e^{\beta(\hbar^{2}k^{2}/2m-\overline{\mu}(\beta,\rho))}-1},
\end{eqnarray*}
since $\forall k\in\Lambda^{*}:\|k\|\leqslant\epsilon,\
k.(x-x')\rightarrow0\ \mathrm{when}\ \epsilon\rightarrow0$. Here
$\overline{\mu}(\beta,\rho):=\lim_{V\uparrow\infty}\overline{\mu}_{\Lambda}(\beta,\rho)$.

Then by virtue of Proposition \ref{thm1g-BEC}
and by the Riemann-Lebesgue Theorem we obtain the result. \qquad \hfill $\Box$ \\

Since Penrose and Onsager \cite{P-O}, the ODLRO is known as the most relevant criterion of
condensation because it is valid with or without interactions between particles.
Here we established that ODLRO is not equivalent to the usual criterion of BEC (macroscopic occupation of the ground state)
but to the generalized BEC. Thus for the PBG a \textit{true} criterion of condensation is the generalized BEC.
{It is natural to suppose (although difficult to check)  the same {for} interacting Bose-gases.}


\subsection{Feynman theory of cycles and existence of long cycles}

\noindent\emph{Feynman concept of cycles}

Here we recall the Feynman concept of cycles introduced in 1953 \cite{F}.
It is related to the Bose statistic and the Feynman-Kac representation for partition functions.

Recall that for the PBG in boxes $\Lambda=L_{1}\times L_{2}\times L_{3}\subset {\mathbb{R}}^3$ with
\textit{periodic} boundary conditions, (i.e. dual vector-spaces $\Lambda^{*}:=\left\{k\in\mathbb{R}^{3}:
k=(2\pi n_{1}/L_{1},2\pi n_{2}/L_{2},2\pi n_{3}/L_{3}),n_{\nu}\in\mathbb{Z}^{1}\right\}$,
see (\ref{dual vector-spaces})) the grand-canonical pressure has the form:
\begin{eqnarray}\label{g-c-pressure}
p_{\Lambda} (\beta, \mu) =\frac{1}{\beta |\Lambda|}\log\left(\Xi_{\Lambda}^{0}(\beta,\mu)\right)
&=&\frac{1}{\beta |\Lambda|} \sum_{k\in\Lambda^{*}}
\ln \left[\left(1 - e^{-\beta(\epsilon_{\Lambda}(k)- \mu)}\right)^{-1}\right]
{}\nonumber\\{}&=&\frac{1}{\beta |\Lambda|}
\sum_{j=1}^{\infty}\ \frac{1}{j}\ e^{j\beta\mu}\ \textrm{Tr}_{\mathcal{H}_{\Lambda}^{(1)}}
\left(e^{-j\beta T_{\Lambda}^{(1)}}\right) \ \ ,
\end{eqnarray}
where $\Xi_{\Lambda}^{0}(\beta,\mu)$ is the PBG grand-canonical partition function and
we used $\textrm{Tr}_{\mathcal{H}_{\Lambda,1}}\left( e^{-j \beta T_{\Lambda}^{(1)}} \right)=
\sum_{\textbf{k}\in\Lambda^{*}} e^{- j \beta \epsilon_{\Lambda}(k)}$.

The Feynman-Kac representation naturally appears from (\ref{g-c-pressure})
if we consider the representation of the trace of
the \textit{Gibbs semigroup}: $\left\{e^{-\beta T_{\Lambda}^{(1)}}\right\}_{\beta>0}$, via its kernel
$K_{\Lambda}(\beta;x,x')=\left(e^{-\beta T_{\Lambda}^{(1)}}\right)(x,x')$:
\begin{equation}\label{TraceNoyau}
\textrm{Tr}_{\mathcal{H}_{\Lambda}^{(1)}}\left(e^{-\beta T_{\Lambda}^{(1)}}\right)
=\int_{\Lambda}dx K_{\Lambda}(\beta;x,x)\ .
\end{equation}

It is known \cite{G} that the kernel $K_{\Lambda}(\beta;x,x')$
can be represented as a path integral over Wiener trajectories starting
at point $x$ and finishing at point $x'$.
Thus $K_{\Lambda}(\beta;x,x)$ can be represent as a Wiener integral over closed trajectories (loops) starting
and finishing at the same point.
The order of the size of the trajectories coincides with the size of the quantum fluctuations $\lambda_{\beta}$,
known as the thermal de Broglie length  \cite{M}.

By virtue of the Gibbs semigroup properties and by expressions (\ref{g-c-pressure}), (\ref{TraceNoyau}), we get:
\begin{equation}\label{g-c-pressureFK}
p_{\Lambda} (\beta, \mu) ={\frac{1}{\beta |\Lambda|}
\sum_{j=1}^{\infty}\ \frac{1}{j}\ e^{j\beta\mu}
\int_{\Lambda}dx_{1}\int_{\Lambda}dx_{2}...\int_{\Lambda}dx_{j}
\ K_{\Lambda}(\beta;x_{1},x_{2})K_{\Lambda}(\beta;x_{2},x_{3})...K_{\Lambda}(\beta;x_{j},x_{1}) \ ,}
\end{equation}
In (\ref{g-c-pressureFK}) each integral over $(x_{1},...,x_{j})\in\Lambda^{j}$
correspond to the impact of the Wiener's loops of the length $j\lambda_{\beta}$ \cite{G}, \cite{M}.

Notice that the total density of particles in the grand-canonical ensemble is given by:
\begin{equation}\label{tot-part-dens0}
\rho_{\Lambda}(\beta, \mu):=\frac{\langle N_{\Lambda}\rangle_{\Lambda}(\beta,\mu)}{V}=
\partial_{\mu}p_{\Lambda} (\beta, \mu) = \frac{1}{V}\sum_{j=1}^{\infty}e^{j\beta\mu} \
\textrm{Tr}_{\mathcal{H}_{\Lambda}^{(1)}} e^{-j\beta T_{\Lambda}^{(1)}} \ \ .
\end{equation}
One can use this representation to identify the \textit{repartition} of the total density (\ref{tot-part-dens0})
over densities of particles involved into loops of the length $j\lambda_{\beta}$:
\begin{equation}\label{rhoj0}
\rho_{\Lambda,j}(\beta, \mu):=
\frac{1}{V}\ e^{j\beta\mu}\ \textrm{Tr}_{\mathcal{H}_{\Lambda}^{(1)}} e^{-j\beta T_{\Lambda}^{(1)}} \ \ .
\end{equation}

The $j$-loops particles density (\ref{rhoj0}) and the representation (\ref{g-c-pressureFK}) are the
key notions for the concept of the \textit{short}/\textit{long} cycles.
Indeed, after the thermodynamical limit one can obtain loops of \textit{finite} sizes or \textit{infinite} sizes,
i.e. one can relate the BEC to appearance of loops of the infinite size as an explanation of the
long-range order and the macroscopic size of the quantum fluctuations \cite{F}, \cite{P-O}, \cite{U}.
{The mathematical basis of the Feynman cycles} approach is related to the boson permutation group,
see \cite{D-M-P},\cite{U}, \cite{S} and \cite{M}. \\

\noindent\emph{BEC and the concept of short/long cycles}

\begin{naming}\label{L-s-cycles}
We say that the representation (\ref{tot-part-dens0}) for the grand-canonical PBG
contains only short cycles if:
\begin{equation}\label{s-cycles}
\rho_{\rm{short}}(\beta,\mu):=
\lim_{M\rightarrow\infty} \left\{\lim_{V\uparrow\infty} \sum_{j=1}^{M}\rho_{\Lambda,j}(\beta,\mu)\right\} =
\rho(\beta, \mu):=\lim_{V\uparrow\infty} \rho_{\Lambda}(\beta, \mu) \ ,
\end{equation}
i.e. it coincides with the total particle density. Since in general the limits in (\ref{s-cycles})
are not interchangeable, we say that for a given total particle density $\rho$ the representation (\ref{tot-part-dens0})
contains macroscopic number of long cycles, if  $\rho>\rho_{\rm{short}}(\beta,\overline{\mu}(\beta,\rho))$,
or equivalently if:
\begin{equation}\label{L-cycles}
\rho_{\rm{long}}(\beta,\rho):=
\lim_{M\rightarrow\infty} \left\{\lim_{V\uparrow\infty}
\sum_{j=M+1}^{\infty}\rho_{\Lambda,j}(\beta,\overline{\mu}_{\Lambda}(\beta,\rho))\right\} > 0 \ .
\end{equation}
\end{naming}

Since Feynman \cite{F} the presence of the non-zero density of the long cycles is usually connected
with the existence of zero-mode BEC, but a rigorous proof of this conjecture for a certain class of models has been
obtained only recently.
There we noticed that even for the PBG type I BEC
the mathematical proof of this connection was not straightforward and appealed to a non-trivial analysis, see \cite{S},
\cite{D-M-P},\cite{U}.

\begin{theorem}\label{Thm1longcycles}
Let us consider a PBG in Casimir boxes $\Lambda$. Then we have:
\begin{eqnarray}
\rho_{\rm{long}}(\beta,\rho)&=&0,\ if\ \rho<\rho_{c}(\beta) \ ,
{}\nonumber\\{}&=&\rho-\rho_{c}(\beta),\ if\ \rho>\rho_{c}(\beta) \ ,
\end{eqnarray}
where $\rho_{c}(\beta)$ is the critical density defined by (\ref{rhoc}).
\end{theorem}

\noindent\emph{Proof}:
By virtue of (\ref{rhoj0}) and (\ref{s-cycles}) we have:
\begin{eqnarray*}
\rho_{\rm{short}}(\beta,\mu)&=&
\lim_{M\rightarrow\infty} \left\{\lim_{V\uparrow\infty} \sum_{j=1}^{M}e^{j\beta\mu}\prod_{\nu=1}^{3}
\frac{1}{V^{\alpha_{\nu}}}
\sum_{n_{\nu}\in\mathbb{Z}^{1}}e^{-j\beta\pi\lambda_{\beta}^{2} (n_{\nu}/V^{\alpha_{\nu}})^{2}} \right\}
{}\\{}&=&\sum_{j=1}^{\infty}e^{j\beta\mu}\prod_{\nu=1}^{3}
\int_{\mathbb{R}}d\xi_{\nu}e^{-j\beta\pi\lambda_{\beta}^{2} \xi_{\nu}^{2}}
=\sum_{j=1}^{\infty}\frac{e^{\beta\mu}}{\lambda_{\beta}^{3}j^{3/2}}=:\frac{g_{3/2}(e^{\beta\mu})}{\lambda_{\beta}^{3}},
\end{eqnarray*}
c.f. definition of $g_{s}(z)$.
By virtue of Proposition \ref{thm1g-BEC}, if $\rho<\rho_{c}(\beta)$ we have $\rho_{\rm{short}}(\beta,\rho)=\rho$
and if $\rho>\rho_{c}(\beta)$ we have $\rho_{\rm{short}}(\beta,\rho)=\rho_{c}(\beta)$.
So, by Definition \ref{L-s-cycles} we conclude the proof. \qquad \hfill $\Box$

Intuitively the size of long cycles for usual BEC (ground state macroscopic occupation)
are of the order of the total particle number ($j=O(N)$).
What happens if the condensate is fragmented,  i.e. of the type II or III ?
Below we apply a scaling approach to study these cases
and we prove in Theorem \ref{thm3} that the order of the size of long cycles is smaller
than the number of particles ($O(N^{\delta}),\ \delta<1$).

\section{Generalized BEC concept: revisited}

In this section we propose a  modification of the concept of generalized BEC, which we call
\emph{scaled} BEC (s-BEC). This implies the corresponding modification of the concept for cycles
(\emph{scaled} short/long cycles denoted s-short/s-long cycles) via similar scaling arguments.
Moreover, we introduce a classification of the s-BEC (type I, II, III) and the \textit{hierarchy} of the
s-long cycles, which distinguishes long-microscopic/macroscopic cycles.

\subsection{Generalized condensation and scaled condensation}

The original van den Berg-Lewis-Pul\'{e} concept of the g-BEC \cite{vdB-L-P} was not explicitly addressed to detect
a \textit{fine} structure of the condensate: \textit{a priori} it does not allow to make the distinction
between generalized BEC of \textit{types} I, II or III. In fact one can do this analysis,
as it was done for the first time in \cite{vdB-L} for the case of the Casimir boxes.

{To make this facet more evident we introduce in this paper a \textit{new} definition of the \textit{generalized} BEC,
which we call a \textit{scaled} BEC (s-BEC).}
Take for simplicity the PBG in Casimir boxes $\Lambda$ with \textit{periodic} boundary conditions,
i.e. with the dual vector-spaces $\Lambda^{*}$ defined by (\ref{dual vector-spaces}),
and with the $k$-modes mean particle densities $\{\rho_{\Lambda}(k)\}_{k\in\Lambda}$
defined by (\ref{meanparticledensities}).
\begin{naming}\label{g-BECvanderBergScale}
We say that for a fixed total density $\rho$ the (perfect) Bose gas manifests a s-BEC in boxes $\Lambda$, if
there exists a positive decreasing function $\eta: V \mapsto {\mathbb{R}}_{+} $, such that
$\lim_{V\uparrow\infty}\eta(V)=0$ and we have:
\begin{equation}\label{s-BEC}
\rho_{\eta}(\beta, \rho) := \liminf_{V\uparrow\infty}
\sum_{\{k\in\Lambda^{*}:\|k\|\leqslant\eta(V)\}}\rho_{\Lambda}(k) > 0 \ .
\end{equation}
\end{naming}
\begin{remark}
Recall that the van den Berg-Lewis-Pul\'{e} definition of g-BEC is formulated as:
\begin{equation}\label{G-BEC}
\rho_{0}(\beta, \rho) := \lim_{\epsilon\downarrow 0}\lim_{V\uparrow\infty}
\sum_{\{k\in\Lambda^{*}:\|k\|\leqslant \epsilon\}}\rho_{\Lambda}(k) > 0 \ .
\end{equation}
Hence, the two Definitions \ref{Defg-BEC} and \ref{g-BECvanderBergScale} are evidently not equivalent.
Moreover, we show that our Definition \ref{g-BECvanderBergScale} allows also to connect a fine mode-structure of
the condensate of types I, II or III with the long-cycles hierarchy, and to show that there is a relation between
the structure of the condensate and the size of cycles.
\end{remark}
The following statement is an evident consequence of Definitions \ref{Defg-BEC} and \ref{g-BECvanderBergScale}:
\begin{lemma}\label{s-G-BEC}
For any function $\eta(V)$ we have:
\begin{equation}\label{s-BEC<G-BEC}
0\leqslant\rho_{\eta}(\beta, \rho) \leqslant \rho_{0}(\beta, \rho) \ .
\end{equation}
\end{lemma}
This lemma means that s-BEC implies g-BEC and (i.e. that if there is no g-BEC,
then there is no s-BEC).

A simple example of application of the s-BEC approach is the possibility to distinguish the
type I, II or III condensations of the PBG in Casimir boxes $\Lambda$ with \textit{periodic}
boundary conditions.
\begin{proposition}\label{eta-dependence-I-II-III}
The rate $\eta(V) = O(1/V^{1/2})$  is an important threshold to refine a
discrimination between different types of g-BEC. If for example, one takes
$\eta_{\delta}(V) = 2\pi/V^{(1/2 -\delta)}$ such that $\delta > 0$,
then we obtain:
\begin{eqnarray*}\label{III-delta}
\rho_{\eta_{\delta}}(\beta,\rho) &=& \rho_{0}(\beta,\rho)  \ , \ \ \ \ \ \ \ \ \ \ \ \ \ \ \rm{for} \
\alpha_{1} \leqslant 1/2  \ ,
\\
\rho_{\eta_{\delta}}(\beta,\rho) &=& 0  \ , \ \ \ \ \ \ \ \ \ \ \ \ \ \ \ \ \ \rm{for} \
\alpha_{1} > 1/2 + \delta  \ ,
\\
\label{III-delta-large}
\rho_{\eta_{\delta}}(\beta,\rho) &=& \rho_{0}(\beta,\rho) \ , \ \ \ \ \ \ \ \ \ \rm{for} \
1/2 + \delta > \alpha_1 > 1/2\ .
\end{eqnarray*}

On the other hand, for $\alpha_1 = 1/2$ and  $\eta_{\Gamma}(V) := 2\pi\Gamma/V^{1/2}$ one gets a modification
of the density of the type II condensation:
\begin{equation*}\label{II-L}
\rho_{\eta_{\Gamma}}(\beta,\rho) = \sum_{|n_{1}|\leqslant \Gamma}\frac{1}{\pi\lambda_{\beta}^{2}n_{1}^{2}+B}
< \rho_{0}(\beta,\rho) \ .
\end{equation*}

For $\alpha_1 > 1/2$ and  $\eta_{\Gamma'}(V) := 2\pi\Gamma'/V^{1-\alpha_{1}}$ one gets a modification of the density of
the type III condensation:
\begin{equation*}\label{III-L}
\rho_{\eta_{\Gamma'}}(\beta,\rho) = \int_{\mathbb{R}^{+}}d\zeta\frac{e^{-C\zeta}}{\zeta^{1/2}\lambda_{\beta}}
\mathrm{erf}(\Gamma'\lambda_{\beta}\sqrt{\zeta\pi})< \rho_{0}(\beta,\rho) \ ,
\end{equation*}
where $\mathrm{erf}(.)$ stands for error function and where $C$ is the unique solution of equation (\ref{EqC}).
\end{proposition}
\noindent\emph{Proof}: By virtue of the proof of the Theorems \ref{thm1}, \ref{thm2} and \ref{thm3},
with the different choices of $\eta(V)$, we can obtain the results. \qquad \hfill $\Box$

Note that the scaling criterion of condensation (Definition \ref{g-BECvanderBergScale})
is more adapted for physical description of BEC because
a judicious choice of the function $\eta(V)$ may give a support of momentum distribution of the condensate.
{It also might be useful for numerical analysis of experimentally observed \textit{fragmented} condensates
\cite{M-H-U-B}.}

{Let $N_{0}$ be a number of particle in the condensate. Then
we say that there is fragmentation of first type if $N_{0}=n_{1}+n_{2}+...+n_{M}$
with $n_{i}=O(N)$ and $M=O(1)$, (where $n_{i}$ are the numbers of particles in the condensate states),
or of the second type if $n_{i}=O(N^{\delta})$ and $M=O(N^{1-\delta})$, $\delta<1$ (such that $N_{0}=O(N)$).}

\subsection{Scaling approach of the short/long cycles}

First we introduce the concepts of \emph{scaled} short/long cycles (s-short/long cycles):
\begin{naming}\label{s-LC}
We say that Bose-gas manifests s-long cycles,
if there exists a positive increasing function of the volume
$\lambda:\mathbb{R}^{+}\rightarrow\mathbb{N}^{+}$, such that $\lim_{V\uparrow\infty}\lambda(V)=\infty$ and:
\begin{equation}
\rho_{\mathrm{long},\lambda}(\beta,\rho):=
\liminf_{V\uparrow\infty}\sum_{j\geqslant\lambda(V)}\rho_{\Lambda,j}(\beta,\overline{\mu}_{\Lambda}(\beta,\rho))>0,
\end{equation}
where $\rho_{\Lambda,j}(\beta,\mu)$ is given by (\ref{rhoj0}).
\end{naming}

The following statement is an evident consequence of Definitions \ref{L-s-cycles} and \ref{s-LC}:
\begin{lemma}
In the particular case of PBG in Casimir boxes we have:
\begin{equation}
0\leqslant\rho_{\mathrm{long},\lambda}(\beta,\rho)\leqslant\rho_{\mathrm{long}}(\beta,\rho),
\end{equation}
for any function $\lambda(V)$.
Here $\rho_{\mathrm{long}}(\beta,\rho)$ is given by (\ref{L-cycles}).
\end{lemma}

This lemma implies that the presence of s-long cycles imply the presence of the long cycles
(c.f. Definition \ref{L-s-cycles}).


A simple example of application of the s-long cycles approach is a possibility to distinguish the
type I, II or III condensations of the PBG in Casimir boxes $\Lambda$ with \textit{periodic}
boundary conditions.

\begin{proposition}\label{propositionlongC}
If $\lambda(V)=V^{\delta}$ then for $\rho\geqslant\rho_{c}(\beta)$ we obtain:
\begin{eqnarray*}
\rho_{\mathrm{long},\lambda}(\beta,\rho)&=&0,\ \mathrm{for}\ \delta>1,
{}\\{}\rho_{\mathrm{long},\lambda}(\beta,\rho)&=&\rho_{0}(\beta,\rho),\ \mathrm{for}\
\alpha_{1}\leqslant1/2,\ 0<\delta<1,
{}\\{}\rho_{\mathrm{long},\lambda}(\beta,\rho)&=&0,\ \mathrm{for}\
\alpha_{1}>1/2,\ 2(1-\alpha_{1})<\delta.
{}\\{}\rho_{\mathrm{long},\lambda}(\beta,\rho)&=&\rho_{0}(\beta,\rho),\ \mathrm{for}\
\alpha_{1}>1/2,\ 0<\delta<2(1-\alpha_{1}),
\end{eqnarray*}
\end{proposition}
\emph{Proof}:
Adapting the proof of the Theorems \ref{thm1}, \ref{thm2} and \ref{thm3}
with the different choices of $\delta$ fot $\lambda(V)=V^{\delta}$ we can obtain the results.
\qquad \hfill $\Box$

\begin{naming}\label{defs-short}
If $j:\mathbb{R}^{+}\rightarrow\mathbb{N}^{+}$ is a bounded positive increasing function
of the volume, i.e. $\lim_{V\uparrow\infty}j(V)<\infty$, then $\rho_{\Lambda,j(V)}
(\beta,\overline{\mu}_{\Lambda}(\beta,\rho))$
is the density of particles in the s-short-cycles of size $j(V)$.
\end{naming}

\begin{naming}\label{defs-long}
If $j:\mathbb{R}^{+}\rightarrow\mathbb{N}^{+}$ is a positive increasing function
of the volume such as $\lim_{V\uparrow\infty}j(V)=\infty$, then $\rho_{\Lambda,j(V)}
(\beta,\overline{\mu}_{\Lambda}(\beta,\rho))$
is the density of particles in the s-long-cycles of size $j(V)$.\\
There is a natural classification of s-long-cycles:

$\bullet$ if $\lim_{V\uparrow\infty}(j(V)/V)=0$, we say that $\rho_{\Lambda,j(V)}(\beta,
\overline{\mu}_{\Lambda}(\beta,\rho))$
is the density of particles in the microscopic-long-cycles of size $j(V)$,

$\bullet$ if $0<\lim_{V\uparrow\infty}(j(V)/V)<\infty$, we say that $\rho_{\Lambda,j(V)}
(\beta,\overline{\mu}_{\Lambda}(\beta,\rho))$
is the density of particles in the macroscopic-cycles of size $j(V)$

$\bullet$ if $\lim_{V\uparrow\infty}(j(V)/V)=\infty$, we say that $\rho_{\Lambda,j(V)}
(\beta,\overline{\mu}_{\Lambda}(\beta,\rho))$
is the density of particles in the large-cycles of size $j(V)$.
\end{naming}
To clarify Definition \ref{defs-long}, we make the following remark:
\begin{remark}\label{Remark1}
We say that if $j(V)$ in (\ref{rhoj0}) is of the order $V^{\alpha}$
if $0<\lim_{V\uparrow\infty}(j(V)/V^{\alpha})<\infty$, for example
$j(V)=xV^{\alpha},\ x>0$. If $\alpha<1$, we are in the first case of the classification in Definition \ref{defs-long}.
{This case is important because it creates a question:}
can we have a macroscopic quantity of particles in the microscopic-long-cycles?
If $\alpha=1$ we are in the second case of the classification of s-long-cycles and if $\alpha>1$ we are in the third case.
Of course, one can take above any increasing function $j(V)$ including e.g. $\ln(V)$.
\end{remark}

\subsection{Hierarchy of s-long cycles}
Definitions \ref{defs-short}, \ref{defs-long} and the Remark \ref{Remark1}
allow to give a natural classification of scaled-long (s-long) cycles.
We call this classification a \emph{hierarchy} of cycles, ordered according their size.

In general there are long cycles of any size:
\begin{naming}\label{s-LCO(g)}
We say that the Bose gas manifests the s-long cycles of the order $\lambda(V)$ where
$\lambda:\mathbb{R}^{+}\rightarrow\mathbb{R}^{+}$
is a positive increasing function of the volume,
if there exist two positive real numbers $x$ and $y$ such as the s-long cycles particle density:
\begin{equation*}
\lim_{V\uparrow\infty}\sum_{j=x\lambda(V)}^{y\lambda(V)}\rho_{\Lambda,j}(\beta,\overline{\mu}_{\Lambda}(\beta,\rho))>0.
\end{equation*}
Then the total density of particles in cycles of size of the order $\lambda(V)$ is:
\begin{equation*}
\rho_{\mathrm{long}}(\beta,\rho|\mathrm{\lambda}):=
\lim_{x\downarrow0;y\uparrow\infty}\lim_{V\uparrow\infty}\sum_{j=x\lambda(V)}^{y\lambda(V)}\rho_{\Lambda,j}
(\beta,\overline{\mu}_{\Lambda}(\beta,\rho)).
\end{equation*}
\end{naming}
We introduce a classification of s-long cycles:
\begin{naming}\label{s-LCmicro}
 - We say that the Bose-gas manifests macroscopic-cycles,
if the Bose gas manifests the s-long cycles of the order $V$.

 - We say that the Bose-gas manifests long-microscopic-cycles,
if the Bose gas manifests the s-long cycles of orders smaller than $V$.
\end{naming}

In the next section we see that in the particular case of the PBG in Casimir boxes,
the classification of s-BEC induces a hierarchy in classification of s-long cycles.
By virtue of Theorem \ref{thm1}, \ref{thm2} and \ref{thm3} we obtain:
\begin{proposition}\label{prop-s-cycles}
If $x$ and $y$ are two positive real numbers, then we have:
\begin{eqnarray*}
\lim_{\Lambda}\sum_{j=x V^{\delta}}^{y V^{\delta}}\rho_{\Lambda,j}(\beta,\overline{\mu}_{\Lambda}(\beta,\rho))
&=&(e^{-xA}-e^{-yA})\rho_{0}(\beta,\rho),\
\mathrm{for}\ \alpha_{1}<1/2,\ \delta=1,
{}\\{}&=&(e^{-x B}-e^{-yB})\rho_{0}(\beta,\rho),\
\mathrm{for}\ \alpha_{1}=1/2,\ \delta=1,
{}\\{}&=&(e^{-x C}-e^{-yC})\rho_{0}(\beta,\rho)\
\mathrm{for}\ \alpha_{1}>1/2,\ \delta=2(1-\alpha_{1}).
\end{eqnarray*}
where $A$ is the unique solution of the equation (\ref{EqA}),
$B$ is the unique solution of the equation (\ref{EqB}) and $C$ is the unique solution of the equation (\ref{EqC}).
\end{proposition}

This proposition gives an illustration of the hierarchy of cycles, in the first case ($\alpha_{1}<1/2$) and second
case ($\alpha_{1}=1/2$) the Bose gas manifests presence of macroscopic-cycles, in the third case ($\alpha_{1}>1/2$),
where the Bose gas manifests presence of long-microscopic-cycles of the order $V^{2(1-\alpha_{1})}$.
A link with the results given by Proposition \ref{thm1g-BEC} we are going to discuss in details in the next Section 3.

\section{Does generalized BEC of the type I, II, III imply a hierarchy of long cycles?}

The aim of this section is to relate a fine structure of g-BEC and s-BEC
(type I, II or III) with hierarchy of long cycles.
Recall that we deal with Casimir boxes $\Lambda=V^{\alpha_{1}}\times V^{\alpha_{2}}\times V^{\alpha_{3}},
\ \alpha_{1}\geqslant\alpha_{2}\geqslant\alpha_{3}>0,\ |\Lambda|=V,\ (\alpha_{1}+\alpha_{2}+\alpha_{3}=1)$.

\subsection{Generalized BEC in the case : $\alpha_{1}<1/2$}

{In this case the geometry is similar to one for the usual cubic box.}
Our main result here is that the g-BEC of type I implies the macroscopic-cycles in the fundamental state.
\begin{theorem}\label{thm1}
If one takes the Casimir boxes $\Lambda$ with $1/2>\alpha_{1}$,
then for a fixed density of particles $\rho>\rho_{c}(\beta)$ the chemical potential is
$\overline{\mu}_{\Lambda}:=\overline{\mu}_{\Lambda}(\beta,\rho)=-A/\beta V+o(1/V)$, with $A>0$.
This implies the s-BEC as well as the g-BEC of type I in the zero mode (ground state)
together only with macroscopic-cycles in this mode.
Here $A$ is the unique solution of equation (\ref{EqA}).
\end{theorem}
\emph{Proof}: Taking into account (\ref{dual vector-spaces}) we denote the family of Casimir boxes by $\Lambda_{I}$
with the corresponding dual space:
\begin{equation}\label{LambdaI}
\Lambda^{*}_{I}:=\left\{k\in\mathbb{R}^{3}:k=(\frac{2\pi n_{1}}{V^{\alpha_{1}}},\frac{2\pi n_{2}}{V^{\alpha_{2}}},
\frac{2\pi n_{3}}{V^{\alpha_{3}}});\ n_{\nu}\in\mathbb{Z}^{1};\ 1/2>\alpha_{1}\right\}.
\end{equation}

\noindent Let $\Lambda^{*}_{0,I}$ be a subset of $\Lambda^{*}_{I}$ defined by:
\begin{equation}\label{lambda0I}
\Lambda^{*}_{0,I}=\{k\in\Lambda^{*}_{I}:\|k\|\leqslant\eta_{I}(V)\},
\end{equation}
where $\eta_{I}(V)=1/V$. Then we have $\Lambda^{*}_{0,I}=\{k=0\}$.

Let us consider the total density of particles:
\begin{equation}\label{rhodec1}
\rho:=\lim_{V\uparrow\infty}\rho_{\Lambda}(\beta,\overline{\mu}_{\Lambda})=
\rho_{\mathrm{short}}(\beta,\rho)+\rho_{\mathrm{long}}(\beta,\rho),
\end{equation}
c.f. (\ref{s-cycles}), (\ref{L-cycles}).

We can decompose the density of particles in long-cycles into two parts defined by:
\begin{equation}\label{rhodec2}
\rho_{\mathrm{long}}(\beta,\rho):=\rho_{\mathrm{long}}(\Lambda^{*}_{I}\backslash\Lambda^{*}_{0,I})
+\rho_{\mathrm{long}}(\Lambda^{*}_{0,I}),
\end{equation}

where $\rho_{\mathrm{long}}(\Lambda^{*}_{I}\backslash\Lambda^{*}_{0,I})$
is the limiting density of particles in long-cycles outside $\Lambda^{*}_{0,I}$:
\begin{equation}\label{longoutsideI}
\rho_{\mathrm{long}}(\Lambda^{*}_{I}\backslash\Lambda^{*}_{0,I}):=\lim_{M\rightarrow\infty}\lim_{V\uparrow\infty}(
\sum_{k\in\Lambda^{*}_{I}\backslash\Lambda^{*}_{0,I}}\sum_{j=M}^{\infty}\rho_{\Lambda,j}(k)),
\end{equation}
where the spectral repartition of particles density in $j$-cycles is:
\begin{equation}
\rho_{\Lambda,j}(k):=\frac{1}{V}e^{j\beta\overline{\mu}_{\Lambda}}e^{-j\beta\epsilon_{\Lambda}(k)},
\end{equation}
and $\overline{\mu}_{\Lambda}:=\overline{\mu}_{\Lambda}(\beta,\mu)$ is the solution of equation
$\rho=\rho_{\Lambda}(\beta,\mu)$.

First we shall estimate the density of particles in long-cycles of $\Lambda^{*}_{I}\backslash\Lambda_{0,I}^{*}$
by (\ref{longoutsideI}) and asymptotic for $\overline{\mu}_{\Lambda}$ we get:
\begin{eqnarray*}\label{EqThmIlongc}
\rho_{\mathrm{long}}(\Lambda^{*}_{I}\backslash\Lambda^{*}_{0,I})&=&
\lim_{M\rightarrow\infty}\lim_{V\uparrow\infty}(\sum_{k\in\Lambda^{*}_{I}\backslash\Lambda^{*}_{0,I}}
\sum_{j=M}^{\infty}\frac{1}{V}e^{j\beta\overline{\mu}_{\Lambda}}e^{-j\beta\epsilon_{\Lambda}(k)}),
{}\nonumber\\{}&=&
\lim_{M\rightarrow\infty}
\sum_{j=M}^{\infty}\frac{1}{(2\pi)^{3}}\int_{\mathbb{R}^{3}}dke^{-j\pi\lambda_{\beta}^{2}k^{2}},
{}\\{}&=&0.
\end{eqnarray*}
Consequently there is no long cycles in $\Lambda^{*}_{I}\backslash\Lambda^{*}_{0,I}$ and
since our last estimate is valid for any $M\rightarrow\infty$ we conclude that there are no s-long cycles
in $\Lambda^{*}_{I}\backslash\Lambda^{*}_{0,I}$ (see Definition \ref{s-LC}).

{Now let us consider the modes in} $\Lambda_{0,I}^{*}$, we would like
to prove that the PBG manifests s-long cycles of the order $O(V)$, i.e. macroscopic-cycles
(see Definition \ref{s-LCmicro}).

Since $\overline{\mu}_{\Lambda}=-A/\beta V+o(1/V)$, with $A>0$ we have:
\begin{eqnarray}\label{rholongMI}
\rho_{\mathrm{long}}(\Lambda_{0,I}^{*}|\mathrm{macro})
&:=&\lim_{x\downarrow0;y\uparrow\infty}\lim_{V\uparrow\infty}\sum_{j=xV}^{yV}
\frac{1}{V}e^{j\beta\overline{\mu}_{\Lambda}}
{}\nonumber\\{}&=&
\lim_{x\downarrow0;y\uparrow\infty}
\lim_{V\uparrow\infty}\frac{(e^{-xA+O(1/V)}-e^{-yA+O(1/V)})}{e^{-\beta\overline{\mu}_{\Lambda}}-1}
{}\nonumber\\{}&=&
\lim_{V\uparrow\infty}
\frac{1}{e^{-\beta\overline{\mu}_{\Lambda}}-1}
=\lim_{V\uparrow\infty}\rho_{\Lambda}(\Lambda_{0,I}^{*}),
\end{eqnarray}
where $\rho_{\Lambda}(\Lambda_{0,I}^{*}):=\sum_{k\in\Lambda_{0,I}}\rho_{\Lambda}(k)$
is the density of particles in $\Lambda_{0,I}^{*}$.

We can easily calculate $\rho_{\Lambda}(\Lambda_{0,I}^{*})$:
\begin{equation}\label{EqrhoI}
\rho_{\Lambda}(\Lambda_{0,I}^{*})=\frac{1}{V}\frac{1}{e^{-\beta\overline{\mu}_{\Lambda}}-1}
=\frac{1}{V}\frac{1}{e^{\beta\frac{A}{\beta V}+o(\frac{1}{V})}-1}=\frac{1}{A}+o(\frac{1}{V}).
\end{equation}
So by virtue of (\ref{rhodec2}), (\ref{EqThmIlongc}) and (\ref{EqrhoI}):
\begin{equation}\label{rholong}
\rho_{\mathrm{long}}(\Lambda_{0,I}^{*}|\mathrm{macro})=\rho_{\mathrm{long}}(\beta,\rho)=\frac{1}{A},
\end{equation}

We know by Theorem \ref{Thm1longcycles} that the density of particles in short-cycles is equal to the critical density.
Consequently by virtue of (\ref{rhodec1}), (\ref{rholong})
we can conclude the proof of the theorem. \qquad \hfill $\Box$

\subsection{Generalized BEC in the case : $\alpha_{1}=1/2$}

{Main result of this sub-section is a theorem about the g-BEC of type II, which is related} to the presence
of macroscopic-cycles in an infinite (in the thermodynamical limit) number of modes.
\begin{theorem}\label{thm2} For the Casimir boxes
with $1/2=\alpha_{1}$ and a fixed particle density $\rho>\rho_{c}(\beta)$
the chemical potential is $\overline{\mu}_{\Lambda}=-B/\beta V+o(1/V)$, with $B>0$.
This implies the s-BEC as well as the g-BEC of type II in an infinite (in thermodynamical limit) number of modes
and simultaneously  macroscopic-cycles only in these modes. Here $B$ is a unique solution of equation (\ref{EqB}).
\end{theorem}
\emph{Proof}: Taking into account (\ref{dual vector-spaces})
we denote the family of Casimir boxes by $\Lambda_{II}$ and the dual space is:
\begin{equation}\label{LambdaII}
\Lambda^{*}_{II}=\left\{k\in\mathbb{R}^{3}:k=(\frac{2\pi n_{1}}{V^{\alpha_{1}}},\frac{2\pi n_{2}}{V^{\alpha_{2}}},
\frac{2\pi n_{3}}{V^{\alpha_{3}}});\ n_{\nu}\in\mathbb{Z}^{1};\ 1/2=\alpha_{1}\right\},
\end{equation}

Let $\Lambda^{*}_{0,II,\Gamma}$ be a subset of $\Lambda^{*}_{II}$
\begin{eqnarray}\label{lambda0II}
\Lambda^{*}_{0,II,\Gamma}=
\left\{k\in\Lambda^{*}_{II}:\|k\|\leqslant \eta_{II}^{\Gamma}(V)\right\},
\end{eqnarray}
where:
\begin{equation}\label{etaII}
\eta_{II}^{\Gamma}(V):=\frac{2\pi \Gamma}{V^{1/2}}, \Gamma\in\mathbb{N}^{*},
\end{equation}

Notice that this set contains the whole value of the condensate as well as particles
involved in the long-cycles for $\Gamma\rightarrow\infty$ after thermodynamical limit.

Again we decompose the density of particles in long cycles into two parts:
\begin{equation}\label{rhodec1II}
\rho_{\mathrm{long}}(\beta,\rho)=
\lim_{\Gamma\uparrow\infty}\rho_{\mathrm{long}}(\Lambda^{*}_{II}\backslash\Lambda^{*}_{0,II,\Gamma})+
\lim_{\Gamma\uparrow\infty}\rho_{\mathrm{long}}(\Lambda^{*}_{0,II,\Gamma}).
\end{equation}
We take the limit $\Gamma\rightarrow\infty$ to have the totality of the condensate in the first part.

The second part of \ref{rhodec1II} is:
\begin{eqnarray}\label{rhodec1IIinterm}
\lim_{\Gamma\uparrow\infty}\rho_{\mathrm{long}}(\Lambda^{*}_{II}\backslash\Lambda^{*}_{0,II,\Gamma})
&=&\lim_{\Gamma\uparrow\infty}\lim_{M\uparrow\infty}\lim_{\Lambda}
\left(\sum_{j=M}^{\infty}\frac{1}{V}e^{-j\beta C/V}
\sum_{\|k\|>2\pi\Gamma/\sqrt{V}}e^{-j\beta\epsilon_{\Lambda}(k)}\right)
{}\nonumber\\{}&=&
\lim_{\Gamma\uparrow\infty}\lim_{M\uparrow\infty}\lim_{\Lambda}
\left(\sum_{j=M}^{\infty}\frac{1}{V}e^{-j\beta C/V}
\sum_{1/V^{1/2-\epsilon}>\|k\|>2\pi\Gamma/\sqrt{V}}e^{-j\beta\epsilon_{\Lambda}(k)}\right)
{}\nonumber\\{}&+&
\lim_{\Gamma\uparrow\infty}\lim_{M\uparrow\infty}\lim_{\Lambda}
\left(\sum_{j=M}^{\infty}\frac{1}{V}e^{-j\beta C/V}\sum_{\|k\|>1/V^{1/2-\epsilon}}e^{-j\beta\epsilon_{\Lambda}(k)}\right),
\end{eqnarray}
with $1/2-\epsilon>\alpha_{2}$. Then we calculate the second term of \ref{rhodec1IIinterm}:
\begin{eqnarray*}
\lim_{\Gamma\uparrow\infty}\lim_{M\uparrow\infty}\lim_{\Lambda}
\left(\sum_{j=M}^{\infty}\frac{1}{V}e^{-j\beta C/V}\sum_{\|k\|>1/V^{1/2-\epsilon}}e^{-j\beta\epsilon_{\Lambda}(k)}\right)
=\lim_{M\uparrow\infty}\sum_{j=M}^{\infty}\frac{1}{j^{3/2}\lambda_{\beta}^{3}}=0.
\end{eqnarray*}
The first term of \ref{rhodec1IIinterm} have an upper bound:
\begin{eqnarray*}
\lim_{\Gamma\uparrow\infty}\lim_{\Lambda}
\left(\sum_{j=1}^{\infty}\frac{1}{V}e^{-j\beta C/V}\sum_{1/V^{1/2-\epsilon}>\|k\|>2\pi\Gamma/\sqrt{V}}
e^{-j\beta\epsilon_{\Lambda}(k)}\right)
=\lim_{\Gamma\uparrow\infty}\sum_{|n_{1}|>\Gamma}\frac{1}{\pi\lambda_{\beta}^{2}n_{1}^{2}+B}=0,
\end{eqnarray*}
consequently the first term of (\ref{rhodec1IIinterm}) is null so
$\lim_{\Gamma\uparrow\infty}\rho_{\mathrm{long}}(\Lambda^{*}_{II}\backslash\Lambda^{*}_{0,II,\Gamma})=0$.

Now let us consider the modes in $\Lambda_{0,II,\Gamma}^{*}$. We would like to apply the same strategy as the proof of the
Theorem \ref{thm1} to prove that the PBG manifests s-long cycles of the order $O(V)$, i.e. macroscopic-cycles
(see Definition \ref{s-LCmicro}).

Since $\overline{\mu}_{\Lambda}=-B/\beta V+o(1/V)$, with $B>0$ we have:
\begin{eqnarray}\label{rholongMII}
\lim_{\Gamma\uparrow\infty}\rho_{\mathrm{long}}(\Lambda^{*}_{0,II,\Gamma}|\mathrm{macro})&:=&
\lim_{\Gamma\uparrow\infty}\lim_{x\downarrow0;y\uparrow\infty}\lim_{V\uparrow\infty}
\frac{1}{V}\sum_{k\in\Lambda_{0,II}^{*}}\sum_{j=xV}^{yV}\rho_{\Lambda,j}(k)
{}\nonumber\\{}&=&\lim_{\Gamma\uparrow\infty}\lim_{x\downarrow0;y\uparrow\infty}
\lim_{V\uparrow\infty}\sum_{k\in\Lambda^{*}_{0,II,\Gamma}}
\frac{(e^{-xB+O(1/V)}-e^{-yB+O(1/V)})}{e^{\beta(\epsilon_{\Lambda}(k)-\overline{\mu}_{\Lambda})}-1}
{}\nonumber\\{}&=&\lim_{\Gamma\uparrow\infty}\lim_{V\uparrow\infty}\sum_{k\in\Lambda^{*}_{0,II,\Gamma}}
\frac{1}{e^{\beta(\epsilon_{\Lambda}(k)-\overline{\mu}_{\Lambda})}-1}
{}\nonumber\\{}&=&\lim_{\Gamma\uparrow\infty}\lim_{V\uparrow\infty}
(\sum_{\textbf{k}\in\Lambda^{*}_{0,II,\Gamma}}\rho_{\Lambda}(\textbf{k})).
\end{eqnarray}
We can easily calculate:
\begin{eqnarray}\label{EqrhoII}
\lim_{\Gamma\uparrow\infty}\lim_{V\uparrow\infty}
(\sum_{\textbf{k}\in\Lambda^{*}_{0,II,\Gamma}}\rho_{\Lambda}(\textbf{k}))
&=&\lim_{\Gamma\uparrow\infty}\lim_{V\uparrow\infty}\sum_{n_{1}=-\Gamma}^{\Gamma}
\frac{1}{V}\frac{1}{e^{\beta(\pi\lambda_{\beta}^{2}n_{1}^{2}/V+B/V+O(1/V))}-1}\nonumber \\
&=&\sum_{n_{1}\in\mathbb{Z}^{1}}\frac{1}{B+\pi\lambda_{\beta}^{2}n_{1}^{2}}.
\end{eqnarray}
So by virtue of (\ref{rhodec1II}), (\ref{rholongMII}) and (\ref{EqrhoII}):
\begin{equation}\label{rholongII}
\lim_{\Gamma\uparrow\infty}\rho_{\mathrm{long}}(\Lambda^{*}_{0,II,\Gamma}|\mathrm{macro})=
\rho_{\mathrm{long}}(\beta,\rho)=\sum_{n_{1}\in\mathbb{Z}^{1}}\frac{1}{B+\pi\lambda_{\beta}^{2}n_{1}^{2}},
\end{equation}
We know by Theorem \ref{Thm1longcycles} that the density of particles in short-cycles is equal to the critical density.
Consequently by virtue of (\ref{rhodec1}), (\ref{rholongII})we can conclude the proof of the theorem.
\qquad \hfill $\Box$

\subsection{Generalized BEC in the case : $\alpha_{1}>1/2$}

Our main result is the theorem about g-BEC of type III due to the presence of
\textit{long-microscopic} cycles in infinite (in the thermodynamical limit) number of modes.
\begin{theorem}\label{thm3}
If one takes the Casimir boxes $\Lambda=V^{\alpha_{1}}\times V^{\alpha_{2}}\times V^{\alpha_{3}}$
with $1/2>\alpha_{1}$, then for a fixed density of particles $\rho>\rho_{c}(\beta)$
the chemical potential is $\overline{\mu}_{\Lambda}(\beta,\rho)=-C/\beta V^{\delta}+o(1/V^{\delta})$,
with $\delta=2(1-\alpha_{1})$ and $C>0$.
This implies the s-BEC as well as the g-BEC of the type III in infinite (in the thermodynamical limit)
number of modes, together with long-microscopic cycles of the order $V^{\delta}$, but only in these modes.
Here $C$ is a unique solution of equation (\ref{EqC}).
\end{theorem}
\emph{Proof}: Taking into account (\ref{dual vector-spaces})
we denote the family of Casimir boxes $\Lambda_{III}$ and the dual spaces are defined by:
\begin{equation}\label{lambdaIII}
\Lambda^{*}_{III}=\left\{k\in\mathbb{R}^{3}:k=(\frac{2\pi n_{1}}{V^{\alpha_{1}}},\frac{2\pi n_{2}}{V^{\alpha_{2}}},
\frac{2\pi n_{3}}{V^{\alpha_{3}}});\ n_{i}\in\mathbb{Z}^{1};\ \alpha_{1}>1/2\right\}\ .
\end{equation}

Let $\Lambda^{*}_{0,III,\Gamma}$ be a subset of $\Lambda^{*}_{III}$:
\begin{eqnarray}\label{lambda0III}
\Lambda^{*}_{0,III,\Gamma}=
\left\{k\in\Lambda^{*}_{III}:\|k\|\leqslant \eta_{III}^{\Gamma}(V)\right\},
\end{eqnarray}
where:
\begin{equation}\label{etaIII}
\eta_{III}^{\Gamma}(V):=\frac{2\pi\Gamma}{V^{\delta/2}},\ \Gamma\in\mathbb{N}^{*},
\end{equation}
here $\delta=2(1-\alpha_{1})<1$.

We show that $\Lambda_{0,III,\Gamma}^{*}$ contains the whole value of the condensate as well as
the particles involved into long cycles in the limit $\Gamma\rightarrow\infty$.

Before to present formal arguments, let us make a remark about \textit{qualitative} difference
between the cases $\alpha_{1}>1/2$ and $\alpha_{1}\leqslant1/2$.
With the definition of $\eta_{III}^{\Gamma}(V)$ (\ref{etaIII}), we see that
the number of states in $\Lambda_{0,III,\Gamma}^{*}$ is of the order $O(V^{2\alpha_{1}-1})$ that goes to infinity,
with increasing volume, there are much more states in the condensate
than in $\Lambda_{0,II,\Gamma}^{*}$ (defined by \ref{lambda0II}).
Heuristically one can say that $\Lambda_{0,III,\Gamma}^{*}$ contains long cycles of sizes of the order $O(V^{\delta})$ in a
number of modes of the order $O(V^{2\alpha_{1}-1})$. Thus the number of particles in these s-long cycles
is of the order $O(V^{\delta})O(V^{2\alpha_{1}-1})=O(V)$, which is macroscopic.
For this reason there is a macroscopic condensate (the order of the number of particles is $O(V)$)
because there is an accumulation of microscopic condensates (the order of the number of particles is
$O(V^{2\alpha_{1}-1})$ which is smaller than $O(V)$)
as well as an accumulation of long-microscopic cycles in $\Lambda_{0,III,\Gamma}^{*}$ at each mode of the condensate.

Again we decompose the density of particles in long cycles into two parts:
\begin{equation}\label{rhodec1III}
\rho_{\mathrm{long}}(\beta,\rho)=
\lim_{\Gamma\uparrow\infty}\rho_{\mathrm{long}}(\Lambda^{*}_{III}\backslash\Lambda^{*}_{0,III,\Gamma})+
\lim_{\Gamma\uparrow\infty}\rho_{\mathrm{long}}(\Lambda^{*}_{0,III,\Gamma}).
\end{equation}
By the same argument as in the proof of Theorem \ref{thm2}, one finds that the first term of \ref{rhodec1III} is null.

Now let us consider the modes in $\Lambda_{0,III,\Gamma}^{*}$. In this case, we would like
to prove that the PBG manifests s-long cycles of the order $O(V^{\delta}),\ \delta=2(1-\alpha_{1})<1$,
i.e. microscopic-cycles (see Definition \ref{s-LCO(g)} and \ref{s-LCmicro}).

Since $\overline{\mu}_{\Lambda}=-C/\beta V^{\delta}+o(1/V^{\delta})$ with $\delta=2(1-\alpha_{1})$ and $C>0$ we obtain:
\begin{eqnarray}\label{rholongmIII}
\lim_{\Gamma\uparrow\infty}\rho_{\mathrm{long}}(\Lambda_{0,III,\Gamma}^{*}|micro)&:=&
\lim_{\Gamma\uparrow\infty}\lim_{x\downarrow0;y\uparrow\infty}
\lim_{V\uparrow\infty}\frac{1}{V}\sum_{k\in\Lambda_{0,III,\Gamma}^{*}}
\sum_{j=xV^{\delta}}^{yV^{\delta}}\frac{1}{V}e^{j\beta\overline{\mu}_{\Lambda}}e^{-j\beta\epsilon_{\Lambda}(k)}
{}\nonumber\\{}&=&
\lim_{\Gamma\uparrow\infty}\lim_{x\downarrow0;y\uparrow\infty}\lim_{V\uparrow\infty}\sum_{k\in\Lambda_{0,III,\Gamma}^{*}}
\frac{(e^{-xC+O(1/V)}-e^{-yC+O(1/V)})}{e^{\beta(\epsilon_{\Lambda}(k)-\overline{\mu}_{\Lambda})}-1}
{}\nonumber\\{}&=&\lim_{\Gamma\uparrow\infty}\lim_{V\uparrow\infty}
(\sum_{\textbf{k}\in\Lambda_{0,III,\Gamma}^{*}}\rho_{\Lambda}(\textbf{k})).
\end{eqnarray}
We can easily calculate:
\begin{eqnarray*}\label{EqrhoIII}
\lim_{\Gamma\uparrow\infty}\lim_{V\uparrow\infty}
(\sum_{\textbf{k}\in\Lambda_{0,III,\Gamma}^{*}}\rho_{\Lambda}(\textbf{k}))
&=&\lim_{\Gamma\uparrow\infty}\lim_{V\uparrow\infty}\left(\frac{1}{V^{\delta}}\frac{1}{V^{2\alpha_{1}-1}}
\sum_{j=1}^{\infty}e^{-(j/V^{\delta})C}\sum_{n_{1}:|n_{1}/V^{2\alpha_1 -1}|\leqslant\Gamma}
e^{-\pi\lambda_{\beta}^{2}(j/V^{\delta})
(n_{1}/V^{2\alpha_{1}-1})^{2}}\right).
\end{eqnarray*}
Since this expression is nothing but the limit of double Darboux-Riemann sums,
in thermodynamical limit we obtain a double integral:
\begin{eqnarray}\label{rholongcalIII}
\lim_{\Gamma\uparrow\infty}\lim_{V\uparrow\infty}
(\sum_{\textbf{k}\in\Lambda_{0,III,\Gamma}^{*}}\rho_{\Lambda}(\textbf{k}))
=\lim_{\Gamma\uparrow\infty}\int_{\mathbb{R}^{+}}d\zeta e^{-\zeta C}
\int_{-\Gamma}^{\Gamma}d\xi e^{-\zeta\pi\lambda_{\beta}^{2}\xi^{2}}
=\int_{\mathbb{R}^{+}}d\zeta \frac{e^{-\zeta C}}{\sqrt{\zeta}\lambda_{\beta}}=\frac{\sqrt{\pi}}{C^{1/2}\lambda_{\beta}}.
\end{eqnarray}
Hence:
\begin{equation}\label{rholongIII}
\lim_{\Gamma\uparrow\infty}\rho_{\mathrm{long}}(\Lambda_{0,III,\Gamma}^{*}|micro)
=\rho_{\mathrm{long}}(\beta,\rho)=\frac{\sqrt{\pi}}{C^{1/2}\lambda_{\beta}} \ ,
\end{equation}
by virtue of (\ref{rhodec1III}), (\ref{rholongmIII}) and (\ref{rholongcalIII}).

We know by Theorem \ref{Thm1longcycles} that the density of particles in short-cycles is equal to the critical density.
Consequently by virtue of (\ref{rhodec1}), (\ref{rholongIII}) we can conclude the proof of the theorem.
\qquad \hfill $\Box$


\section{Does generalized BEC I, II, III imply a hierarchy of ODLRO?}


In the Introduction we presented three concepts related to the BEC: g-BEC, long cycles and ODLRO.
We present in Sections 2 and 3 two new concepts: \emph{scaled} BEC (s-BEC) and \emph{scaled} short/long cycles
(s-short/long cycles) associated with g-BEC and short/long cycles. Thus it seems consistent to introduce a concept
of \emph{scaled} ODLRO (s-ODLRO) via our scaling approach to study the hierarchy of ODLRO.

This part gives a physical meaning of the scaling approach because it allows
the study of the coherence of the condensate at large scale.
We show that for very anisotropic cases ($\alpha_{1}>1/2$) the coherence length (maximal length of correlation)
is not equal to the size of the box (see Theorem \ref{TmODLRO}).

\subsection{Scaling approach to ODLRO}

Recall that the generalized criterion of ODLRO is that there is ODLRO if and only if there is g-BEC,
see Theorem \ref{ThmODLRO}.
The standard definition of ODLRO is formulated in Definition \ref{DefODLRO}:
\begin{equation*}\label{EqDefODLRO}
\sigma(\beta,\rho):=\lim_{\|x-x'\|\uparrow\infty}\sigma(\beta,\rho;x,x'),
\end{equation*}
where $\sigma(\beta,\rho;x,x')$ is the two-point correlation function
between two points $x$ and $x'$ after thermodynamical limit.
Notice that this definition can not be satisfactory when we use the definition of s-BEC,
since we do not precise yet what are the scales of large correlations.
It seems to be interesting to take thermodynamical limit at the same time
as we take the two points $x$ and $x'$ at increasing distance.

A natural question is whether we are able to detect different types of s-BEC (as well as g-BEC) with
the help of a generalized criterion of ODLRO based on our scaling approach? We call it a \emph{scaled} ODLRO (s-ODLRO).

\begin{naming}\label{Def-s-ODLRO}
The PBG manifests a s-ODLRO
if there exists a vector-valued function of volume $X:V\mapsto X(V)\in\Lambda$ such that
$\lim_{V\uparrow\infty}|X_{\nu}(V)|=\infty,\ \nu=1,2,3$ and:
\begin{equation}\label{sigmaX}
\sigma_{X}(\beta,\rho):=\lim_{V\uparrow\infty}(\sigma_{\Lambda,X})(V)>0,
\end{equation}
where $(\sigma_{\Lambda,X})(V)$ is the two-point scaled-correlation function (two-point s-correlation function)
for $x(V),x'(V)\in\Lambda$, see (\ref{EqDefCorrel}):
\begin{equation}\label{sigmaX(V)}
(\sigma_{\Lambda,X})(V):=\sigma_{\Lambda}(\beta,\rho;x(V)-x'(V))=\sum_{k\in\Lambda^{*}}\rho_{\Lambda}
(k)e^{ik\cdot X(V)},
\end{equation}
here $X(V)=(x-x')(V)\in\Lambda$.
\end{naming}

\begin{remark}\label{RqsODLRO}
By (\ref{dual vector-spaces}) and (\ref{meanparticledensities}) one can write (\ref{EqDefCorrel}) like:
\begin{equation}\label{RkODLRO}
\sigma(\beta,\rho;x,x')=\sum_{j=1}^{\infty}e^{j\beta\overline{\mu}_{\Lambda}}\prod_{\nu=1}^{3}
\theta_{3}(\frac{\pi}{V^{\alpha_{\nu}}}(x_{\nu}-x_{\nu}^{'}),e^{-j\pi\frac{\lambda_{\beta}^{2}}{ V^{2\alpha_{\nu}}}})\ ,
\end{equation}
where $\theta_{3}(u,q):=\sum_{n\in\mathbb{Z}^{1}}q^{n^{2}}e^{2inu}$ is the elliptic theta-function.
\end{remark}

This implies the following proposition:
\begin{proposition}\label{propsODLRO}
By (\ref{sigmaX(V)}) the two-point correlation function as well as the two-point s-correlation function
(see Definition \ref{Def-s-ODLRO}) are  non-negative symmetric and $L_{\nu}$-periodic functions of
$x_{\nu}-x'_{\nu},\ \nu=1,2,3$ on $\mathbb{R}$, and decreasing/increasing on
$\left[nL_{\nu},nL_{\nu}+L_{\nu}/2\right]\subset\mathbb{R}^{+},\ n\in\mathbb{N}$,
respectively on $\left[nL_{\nu}+L_{\nu}/2,(n+1)L_{\nu}\right]\subset\mathbb{R}^{+},\ n\in\mathbb{N}$
(i.e. monotone on the semi-periods).
\end{proposition}
\emph{Proof}: These  properties follow from the properties of the elliptic theta-function \cite{A-S}.
\qquad \hfill $\Box$

The following statement {is a direct consequence} of Definitions \ref{DefODLRO} and Remark \ref{Def-s-ODLRO}:
\begin{lemma}\label{l1}
For any vector-valued $X(V)$ we have:
\begin{equation}
0\leqslant\sigma_{X}(\beta,\rho)\leqslant\sigma(\beta,\rho)\ .
\end{equation}
\end{lemma}
This Lemma means that the s-ODLRO implies standard ODLRO.

\subsection{Hierarchy and anisotropy of ODLRO, coherence of the condensate}

Here we use Definition \ref{Def-s-ODLRO} to analyze the s-BEC and the s-long cycles in the Casimir boxes.
Notice that the usual criterion of ODLRO is such that we have no indication of the scale of long correlations
because we study their correlations after thermodynamical limit.

We introduce a classification of the s-OLDRO which is formally defined by:
\begin{naming}\label{defhieraODLROmacro}
The PBG manifests the macroscopic-ODLRO in the direction $x_{\nu}$, if
there exist a vector $X(V)=(X_{1}(V),X_{2}(V),X_{3}(V))\in\Lambda$ such that:
$\lim_{V\uparrow\infty}|X_{\nu}(V)|/V^{\alpha_{\nu}}>0$ and $\sigma_{X}(\beta,\rho)>0$.
\end{naming}

\begin{naming}\label{defhieraODLROmicro}
If the PBG does not manifests the macroscopic-ODLRO in the direction $x_{\nu}$
although the PBG manifests the s-ODLRO, then it manifests the microscopic-ODLRO in the direction $x_{\nu}$.
\end{naming}

With the periodic boundary conditions the system
is homogeneous and so there is no localization of the condensate in the space contrary to the case of the
Dirichlet boundary conditions. However the coherence length of the condensate could be studied
{on the basis of  the precedent Definitions \ref{defhieraODLROmacro} and \ref{defhieraODLROmicro}.}

\begin{theorem}\label{TmODLRO}
Let us consider the grand-canonical PBG in Casimir boxes $\Lambda=V^{\alpha_{1}}\times V^{\alpha_{2}}\times V^{\alpha_{3}}$
with Dirichlet boundary conditions and a fixed density of particles $\rho$.
Let $X:V\in\mathbb{R}^{+}\mapsto X(V)=(X_{1}(V),X_{2}(V),X_{3}(V))\in\Lambda,
\ \lim_{V\uparrow\infty}X_{\nu}(V)=\infty,\ 0<X_{\nu}(V)\leqslant V^{\alpha_{\nu}}/2,\ \nu=1,2,3$.
Then we have the following results concerning the s-ODLRO, see (\ref{sigmaX}):
\begin{eqnarray}\label{eqTmSODLRO0}
\sigma_{X}(\beta,\rho)=0,\ \mathrm{for}\ \rho<\rho_{c}(\beta),
\end{eqnarray}
Whereas for $\rho>\rho_{c}(\beta)$ we get:

for $\alpha_{1}<1/2$:
\begin{eqnarray}
\sigma_{X}(\beta,\rho)&=&\rho_{0}(\beta)\ ,
\end{eqnarray}

for $\alpha_{1}=1/2$:
\begin{eqnarray}
\sigma_{X}(\beta,\rho)&=&\rho_{0}(\beta),
\ \mathrm{for}\ \lim_{V\uparrow\infty}(X_{1}(V)/V^{\alpha_{1}})=0\ ,
{}\\{}&=&\sum_{n_{1}\in\mathbb{Z}^{1}}
\frac{\cos{2\pi n_{1}x}}{\pi\lambda_{\beta}^{2}n_{1}^{2}+B}<\rho_{0}(\beta),
\ \mathrm{for}\ X_{1}(V)=xV^{\alpha_{1}}/2,\ 0<x<1,\ ,
\end{eqnarray}

for $\alpha_{1}>1/2$:
\begin{eqnarray}
\sigma_{X}(\beta,\rho)&=&\rho_{0}(\beta),\ \mathrm{for} \lim_{V\uparrow\infty}(X_{1}(V)/V^{\delta})=0,
\ \delta=2(1-\alpha_{1})\ ,
{}\\{}&=&\rho_{0}(\beta)e^{-2x \sqrt{\pi c}/\lambda_{\beta}}<\rho_{0}(\beta)
,\ \mathrm{for}\ X_{1}(V)=xV^{\delta/2},\ x>0\ ,
{}\\{}&=&0,\ \mathrm{for}\ \lim_{V\uparrow\infty}(X_{1}(V)/V^{\delta/2})=0\ .
\end{eqnarray}
\end{theorem}
\emph{Proof}: {To ensure a monotonous decreasing of the correlation functions for the case of periodic
boundary conditions,}
we choose $0<X_{\nu}\leqslant\frac{1}{2}V^{\alpha_{\nu}},\ \nu=1,2,3$, see Proposition \ref{propsODLRO}.

The first step of the proof is to study the case $\rho<\rho_{c}(\beta)$:

Since $\sigma(\beta,\rho)=0$ (Theorem \ref{ThmODLRO}),
by Lemma \ref{l1} we get $\sigma_{X}(\beta,\rho)=0$ for any vector $X(V)\in\Lambda$.

The second step is to study the case $\rho>\rho_{c}(\beta)$:

For $\alpha_{1}<1/2$ by Definition \ref{Def-s-ODLRO} we have:
\begin{eqnarray}\label{Eq1}
\sigma_{X}(\beta,\rho)&=&\lim_{V\uparrow\infty}\left(\sum_{k\in\Lambda_{0,I}^{*}}\rho_{\Lambda}(k)e^{ik.X(V)}\right)
+\lim_{V\uparrow\infty}\left(\sum_{k\in\Lambda_{I}^{*}\backslash\Lambda_{0,I}^{*}}\rho_{\Lambda}(k)e^{ik.X(V)}\right)
{}\nonumber\\{}&=&\lim_{V\uparrow\infty}\left(\frac{1}{V}\sum_{j=1}^{\infty}e^{-Aj/V}\right)
+\lim_{V\uparrow\infty}\left(\sum_{k\in\Lambda_{I}^{*}\backslash\Lambda_{0,I}^{*}}\rho_{\Lambda}(k)e^{ik.X(V)}\right)
\end{eqnarray}
where $\Lambda_{I}^{*}$ is the dual vector space given by equation (\ref{LambdaI}) and
$\Lambda_{0,I}^{*}=\{k=0\}$ is the subset corresponding to the condensate defined by (\ref{lambda0I}).
The first term of (\ref{Eq1}) is equal to $\rho_{0}(\beta,\rho)$ by virtue of Theorem \ref{thm1},
so given that $\sigma(\beta,\rho)=\rho_{0}(\beta,\rho)$ (Theorem \ref{ThmODLRO}) and by Lemma \ref{l1}
the second term of (\ref{Eq1}) have to be null and we get the result.

For $\alpha_{1}=1/2$ by Definition \ref{Def-s-ODLRO} we obtain:
\begin{eqnarray}\label{Eq2}
&&\sigma_{X}(\beta,\rho)=\lim_{\Gamma\uparrow\infty}\lim_{V\uparrow\infty}\left(\sum_{k\in\Lambda_{0,II,\Gamma}^{*}}
\rho_{\Lambda}(k)e^{ik.X(V)}\right)
+\lim_{\Gamma\uparrow\infty}\lim_{V\uparrow\infty}\left(\sum_{k\in\Lambda_{II}^{*}\backslash\Lambda_{0,II,\Gamma}^{*}}
\rho_{\Lambda}(k)e^{ik.X(V)}\right)
{}\nonumber\\{}&=&\lim_{\Gamma\uparrow\infty}\lim_{V\uparrow\infty}\left(\frac{1}{V}\sum_{j=1}^{\infty}e^{-Bj/V}
\sum_{n_{1}=-\Gamma}^{\Gamma}e^{-\pi\lambda_{\beta}^{2}n_{1}^{2}(j/V)}e^{2\pi i X_{1}(V)(n_{1}/V)}\right)
{}\nonumber\\{}&+&\lim_{\Gamma\uparrow\infty}\lim_{V\uparrow\infty}
\left(\sum_{k\in\Lambda_{II}^{*}\backslash\Lambda_{0,II,\Gamma}^{*}}\rho_{\Lambda}(k)e^{ik.X(V)}\right)
\end{eqnarray}
where $\Lambda_{II}^{*}$ is the dual vector space given by equation (\ref{LambdaII}) and
$\Lambda_{0,II,\Gamma}^{*}$ is the subset corresponding to the condensate defined by (\ref{lambda0II}).

If $\lim_{V\uparrow\infty}(X_{1}(V)/V^{\alpha_{1}})=0$,
the first term in the right-hand side of (\ref{Eq2}) is equal to $\rho_{0}(\beta,\rho)$ by virtue of Theorem \ref{thm2}.
Given that $\sigma(\beta,\rho)=\rho_{0}(\beta,\rho)$ (Theorem \ref{ThmODLRO}) and by Lemma \ref{l1},
the second term in (\ref{Eq2}) is null and we obtain the result.

Let $\lim_{V\uparrow\infty}(X_{1}(V)/V^{\alpha_{1}})=x,\ 0<x\leqslant1/2$.
Since the sum inside the limit in the first term of the right-hand side of (\ref{Eq2})
is a Darboux-Riemann sum, one get:
\begin{equation*}
\int_{\mathbb{R}^{+}}d\chi e^{-B\chi}\sum_{n_{1}\in\mathbb{Z}^{1}}e^{-\pi\lambda_{\beta}^{2}n_{1}^{2}\chi}e^{2\pi i xn_{1}}
=\sum_{n_{1}\in\mathbb{Z}^{1}}\frac{\cos{2\pi n_{1}x}}{\pi\lambda_{\beta}^{2}n_{1}^{2}+B}.
\end{equation*}
The second term in the right hand side of (\ref{Eq2}) is null because the phase implies that it is
smaller than the density of particles in $\Lambda^{*}_{II}\backslash\Lambda^{*}_{II,0,\Gamma}$
which is null in the limit $\Gamma\rightarrow\infty$.

For $\alpha_{1}>1/2$ by Definition \ref{Def-s-ODLRO} and by virtue of (\ref{EqrhoIII})  we have:
\begin{eqnarray}\label{Eq3}
\sigma_{X}(\beta,\rho)&=&\lim_{\Gamma\uparrow\infty}\lim_{V\uparrow\infty}\left(\sum_{k\in\Lambda_{0,III,\Gamma}^{*}}
\rho_{\Lambda}(k)e^{ik.X(V)}\right)
+\lim_{\Gamma\uparrow\infty}\lim_{V\uparrow\infty}\left(\sum_{k\in\Lambda_{III}^{*}\backslash\Lambda_{0,III,\Gamma}^{*}}
\rho_{\Lambda}(k)e^{ik.X(V)}\right)
{}\nonumber\\{}&=&
\lim_{\Gamma\uparrow\infty}\lim_{V\uparrow\infty}\left(\frac{1}{V^{\delta}}\frac{1}{V^{2\alpha_{1}-1}}
\sum_{j=1}^{\infty}e^{-Cj/V^{\delta}}\sum_{n_{1}:|n_{1}/V^{2\alpha_1 -1}|\leqslant\Gamma}
e^{-\pi\lambda_{\beta}^{2}(j/V^{\delta})(n_{1}/V^{2\alpha_{1}-1})^{2}}e^{2\pi i X_{1}(V)n_{1}/V^{\alpha_{1}}}\right)
{}\nonumber\\{}&+&\lim_{\Gamma\uparrow\infty}\lim_{V\uparrow\infty}
\left(\sum_{k\in\Lambda_{III}^{*}\backslash\Lambda_{0,III,\Gamma}^{*}}\rho_{\Lambda}(k)e^{ik.X(V)}\right)
\end{eqnarray}
where $\Lambda_{III}^{*}$ is the dual vector space given by equation (\ref{lambdaIII}) and
$\Lambda_{0,III}^{*}$ is the subset corresponding to the condensate defined by (\ref{lambda0III}).

If $\lim_{V\uparrow\infty}(X_{1}(V)/V^{\delta/2})=0$,
the right-hand side of (\ref{Eq3}) is equal to $\rho_{0}(\beta,\rho)$ by virtue of Theorem \ref{thm3}.
Since $\sigma(\beta,\rho)=\rho_{0}(\beta,\rho)$ (Theorem \ref{ThmODLRO}), by Lemma \ref{l1}
the second term of (\ref{Eq3}) have to be null and we get the result.

Let $\lim_{V\uparrow\infty}(X_{1}(V)/V^{\delta/2})=x,\ x>0,\ \delta=2(1-\alpha_{1})$.
Then the sum inside the limit in the first term of (\ref{Eq3}) is a double Darboux-Riemann sum, which implies:
\begin{equation*}
\int_{\mathbb{R}^{+}}d\xi e^{-C\xi}\int_{\chi\in\mathbb{R}}d\chi e^{-\pi\lambda_{\beta}^{2}\xi\chi^{2}}e^{2\pi ix\chi}
=\frac{\sqrt{\pi}}{\lambda_{\beta}\sqrt{C}}e^{-2x \sqrt{\pi c}/\lambda_{\beta}}.
\end{equation*}

By the same argument as in the case $\alpha_{1}=1/2$ the second part of (\ref{Eq3}) is null
(the precedent expression is a decreasing function of $x$ for $x>0$) and thus we obtain the result.

Let $\lim_{V\uparrow\infty}(X_{1}(V)/V^{\delta/2})=\infty$.
Since the correlation function is a decreasing function for $0<X_{\nu}\leqslant V^{\alpha_{\nu}}/2$
(see Proposition \ref{propsODLRO}), it is uniformly bounded by the above estimate with $X(V)=xV^{\delta},\ x>0$:
\begin{eqnarray*}
\int_{\mathbb{R}^{+}}d\xi e^{-C\xi}\int_{\chi\in\mathbb{R}}d\chi e^{-\pi\lambda_{\beta}^{2}\xi\chi^{2}}e^{2\pi i x\chi}
+\lim_{\Gamma\uparrow\infty}\lim_{V\uparrow\infty}
\left(\sum_{k\in\Lambda_{III}^{*}\backslash\Lambda_{0,III,\Gamma}^{*}}\rho_{\Lambda}(k)e^{ik.X(V)}\right).
\end{eqnarray*}
When $x$ tends to infinity, the first part goes to zero (by the Riemann-Lesbegue theorem) then the precedent arguments
show that the second part is also null. This concludes the proof. \qquad \hfill $\Box$

We give here a classification of the s-ODLRO for three cases of Casimir boxes:
\begin{theorem}\label{ThmhieraODLRO}
If one takes the Casimir boxes with $\alpha\leqslant1/2$ then for a fixed density $\rho>\rho_{c}(\beta)$
the PBG manifests macroscopic-ODLRO in three directions.
If $\alpha>1/2$ then for a fixed density $\rho>\rho_{c}(\beta)$
the PBG manifests microscopic-ODLRO in direction $x_{\nu}$ and macroscopic-ODLRO
in other directions.
\end{theorem}
\emph{Proof}: Definitions \ref{defhieraODLROmacro}, \ref{defhieraODLROmicro}
and Theorem \ref{TmODLRO} give the proof of the theorem. \qquad \hfill $\Box$

It is remarkable that for type I and II g-BEC in Casimir boxes corresponding to
$\alpha_{1}\leqslant1/2$ the condensate is spatially macroscopic whereas for the case $\alpha_{1}>1/2$
the condensate is spatially macroscopic in two directions but microscopic in the most anisotropic direction $x_{1}$.
It is naturally to guess that there is a link between the size of s-long cycles and coherence length
of the condensate. We can see this explicitly in \cite{U}
where the competition between the size of correlation $X$ and the size of cycle $j$ indicates
that the coherence length is of the order of the square root of the size
of the s-long cycles (e.g. $V^{\delta/2},\ \delta=2(1-\alpha_{1})$
in the case of the PBG in Casimir boxes with $\alpha_{1}>1/2$).

\section{Concluding remarks}

\noindent\emph{Some technical remarks}

In this paper we introduce a new concept of BEC which is called the scaled BEC (s-BEC)
to adapt the London scaling approach to the problem of g-BEC.
It implies the van den Berg-Lewis-Pul\'e classification of BEC in three types (I,II,III)
illustrated for the particular case of the PBG in Casimir boxes.
This is a first formal step necessary before to study more carefully the different case of g-BEC
for the PBG in Casimir boxes. One can see this by virtue of Proposition and \ref{eta-dependence-I-II-III}.

The fundamental question that we study in this paper is the relation of different types
of g-BEC (I,II,III) with the long cycles and with the ODLRO. Our results concerning the PBG in Casimir boxes
can be summarized as follows:

 - we introduced new concept of short/long cycles called scaled short/long cycles (s-short/long cycles,
 see Definition \ref{s-LC}) to distinguish
 different types of g-BEC, see Theorems \ref{thm1}, \ref{thm2} \ref{thm3} and Remark \ref{Remark1}
 and Proposition \ref{prop-s-cycles}.
 This paper is based on the estimation of the size of s-long cycles in the condensate, see Definitions \ref{defs-short}
 at \ref{s-LCO(g)}. If the size of s-long cycles is macroscopic, then the g-BEC is of the type I or II but if the
 s-long cycles are microscopic, then the g-BEC is of type III.

 - we introduced new concept of ODLRO, called scaled ODLRO (s-ODLRO, see Definition \ref{Def-s-ODLRO}), to distinguish
 the different types of g-BEC, see Theorems \ref{TmODLRO}, \ref{ThmhieraODLRO}.
 Our arguments are based on the estimate of the coherence length of the condensate, see Definitions \ref{defhieraODLROmacro},
 \ref{defhieraODLROmicro}.
 If the coherence length is macroscopic in the three directions, then the g-BEC is of type I or II,
 and if in one of the three dimensions the coherence length is microscopic, then the g-BEC is of type III.

{It is clear that the proof} of the Theorem \ref{thm1}, \ref{thm2} and \ref{thm3}
is based on an analysis of geometric series easily done via s-long cycles. For this reason we can say that s-long cycles
is a well adapt technique to study classification of BEC. Another reason is that the concept of cycles is
independent of the representation of the gas (Feynman-Kac versus spectral representation). {Therefore the cycles seem to
be useful to study generalized BEC for an interacting Bose gas.}

For simplicity we consider here the PBG with periodic boundary conditions
but one can adapt present paper to Dirichlet or Neumann boundary conditions for which we can characterize the
geometric form of the condensate cloud
via the concept of scaled local density. In these cases the results concerning the hierarchies of cycles and
ODLRO do not change.
However, if one takes attractive boundary conditions,
see \cite{V-V-Z}, we guess that the result should be different, since in this case the condensate
is localized in two modes and it is not homogeneous.
One can suppose that the s-long cycles are macroscopic but the most interesting is the coherence length
of the condensate and its geometric form.

The last technical remark concerns the validity of the results in the canonical ensemble.
It is known \cite{P-Z} that in the canonical ensemble for the PBG in Casimir boxes
there is the same type of generalized BEC as in the grand-canonical ensemble for the equivalent geometry.
Thus we believe that the principle conclusions concerning the hierarchy of long cycles and of ODLRO
should not change. However given that the amount of condensate in individual states are different \cite{P-Z}
from the grand-canonical case and some results (formulae in the Propositions \ref{eta-dependence-I-II-III}
and \ref{prop-s-cycles} and Theorem \ref{TmODLRO}) should be different.  \\

\noindent\emph{Conceptual and physical remarks} \\

First notice that ODLRO, long cycles and generalized BEC are equivalent criterions for PBG in Casimir boxes.
Consequently generalized BEC is more relevant than usual BEC (macroscopical occupation of ground state)
because it contains all cases of BEC (fragmented or not).

In this article we present a scaling approach of Bose-Einstein condensation. It allows classification
of different types of condensate via scaling size of long cycles in relation with scaling size of large correlations.
The interest of the study of long cycles using scaling approach is the knowledge on coherence properties
of condensate by virtue of the rule of Bose statistics in the two points correlation function.
Heuristically the order of the size of large correlation is the square root of the order of the size of long cycles.

Scaling approach should be useful for interpretation of thermodynamical results for large particles number systems.
It is interesting because one can study the effect of the geometry of the box
on geometry and coherence properties of the condensate.
In this paper, we show that a condensate is like a finite or infinite number of macroscopic particle (type I or II)
or infinite number of microscopic particles (type III, analogous to quasi-condensate \cite{M-H-U-B}, \cite{P-S-W})
formed by "closed polymers chain" (cycles) of macroscopic or microscopic size related to coherence length
(square root of size of cycles).
Physically it could be interesting to study the correlation function of a condensate in an harmonic trap with pulsations
$\omega_{x}, \omega_{y}, \omega_{z}$ using our scaling approach (e.g. for very anisotropic traps). \\

\noindent\emph{Perspectives} \\

{In the present paper we the choice of Casimir boxes serves to illustrate our concepts. But we can use
the \emph{van den Berg} boxes, which are a generalization of Casimir ones}
($\Lambda_{L}=L_{1}(L)\times L_{2}(L)\times L_{3}(L)$ with $V_{L}:=|\Lambda_{L}|=L_{1}(L)L_{2}(L)L_{3}(L)$
where $L_{i}(L)$ are functions
of a parameter $L$ such as $\lim_{L\rightarrow\infty}L_{i}(L)=\infty$).
These boxes are {very interesting to study since} with particular choice
of the functions $L_{i}(L)$,  e.g. (see\cite{vdB}) $L_{1}(L)=L_{2}(L)=e^{L},L_{3}(L)=L$ with
Dirichlet boundary conditions,
for $\rho_{c}(\beta)<\rho<\rho_{m}(\beta)$ the g-BEC is of type III
and for $\rho>\rho_{m}(\beta)$ it seems there is the \emph{coexistence} of g-BEC of type I and type III,
where $\rho_{c}(\beta)$ is the critical density defined by (\ref{rhoc}) and $\rho_{m}(\beta)$ is
a critical density defined in this particular case by $\rho_{m}(\beta)=\rho_{c}(\beta)+1/\beta\pi$.
In fact $\rho_{c}(\beta)$ separates regimes condensate - non condensate and $\rho_{m}(\beta)$ separates
the different types of generalized BEC. {This seems to be analogous to }quasi-condensate/condensate transition
\cite{P-S-W}. It is natural now to study this curious phenomena using our approach,
which will be {a subject of another paper.}  \\

Whether the scaling approach for the interacting gas problem is still relevant?
One can see that the concept of g-BEC is well formulated for the PBG
but not for the interacting Bose gas. Then how to study the classification of BEC for the interacting Bose gas
if we have no more indications about the spectral properties of the gas?
This problem could be solved using our scaling approach (of long cycles or ODLRO).
Thus the next step will be the application of these new methods for some models of interacting Bose gas
(especially for weakly interacting Bose gas at first time, take e.g. an interaction
$U=\sum_{k\in\Lambda^{*}}g N_{k}^{2}/2V,\ g>0$, see \cite{Br-Z})).
For strongly interacting bosons in \textit{quantum crystals}, there is a formation of infinite cycles \cite{U2}.
It could be interesting to study them using scaling approach of long cycles.

Another interesting problem is the characterization of quantum fluctuations
in many body systems. Scaling approach could be also relevant to study finite
size scaling effect \cite{B-D-T} appearing in some systems,
like Casimir effect in quantum liquids \cite{M-Z}, \cite{Z-al}. \\

\textbf{Acknowledgments} \\

I would like to thank my promotor Valentin Zagrebnov, {who introduced me into this subject, for encouragements,
helpful remarks and numerous discussions. I benefited a lot from the referees' constructive criticism and
suggestions which improved the first version of the manuscript.}


\end{document}